\documentclass{article}
\usepackage{amssymb,amsfonts,amsmath}
\usepackage{cite,enumerate,float}
\usepackage{color}
\usepackage{tikz}
\usetikzlibrary{arrows,snakes,backgrounds}
\DeclareGraphicsExtensions{.pdf,.png,.jpg}
\usepackage{multirow}
\usepackage{pgfplots}
\pgfplotsset{compat=1.9}
\usepackage{amsmath, amsfonts, amssymb, amsthm, mathtools}
\usepackage{rotating}
\usepackage{ytableau}
 \ytableausetup{mathmode, boxsize=0.4 em}
\usepackage{braket}
\usepackage{float}
\usepackage{amsmath}
\usepackage{tikz}
\usetikzlibrary{calc,matrix}
\usepackage{wrapfig}
\usepackage{graphicx}
\graphicspath{{pictures/}}
\usepackage{hyperref}
\DeclareGraphicsExtensions{.pdf,.png,.jpg}
\usepackage{array}
\usepackage[T2A]{fontenc}
\usepackage[utf8]{inputenc}
\def\be{\begin{eqnarray}}
\def\ee{\end{eqnarray}}
\def\nn{\nonumber}
\def\p{\partial}
\def\tr{{\rm tr}\,}

\definecolor{red}{rgb}{1,0,0}
\definecolor{orange}{rgb}{1,0.5,0}
\definecolor{violet}{rgb}{0.7,0,1}

\hoffset= - 1.0in

\begin{document}


\hfill MIPT/TH-09/25

\hfill IITP/TH-09/25

\hfill ITEP/TH-11/25

\bigskip

\centerline{\Large{Itzykson-Zuber correlators from character expansion
}}

\bigskip
\renewcommand{\thefootnote}{\fnsymbol{footnote}}
\centerline{\textbf{Alexei Morozov$^{1,2,3}$\footnote[2]{morozov@itep.ru} \ and\ \   Hasib Sifat$^1$\footnote[3]{ sifat.ma@phystech.edu}}}

\bigskip

\centerline{\it $^1$MIPT, 141701, Dolgoprudny, Russia}
\centerline{\it $^2$IITP RAS, 127051, Moscow, Russia}
\centerline{\it $^3$ITEP, Moscow, Russia}

\bigskip

\centerline{ABSTRACT}

\bigskip

{\footnotesize
We demonstrate the consistency of character expansion for the Itzykson-Zuber (IZ) model in terms of Schur polynomials with the old formulas for pair correlators with the IZ measure. An essential new feature of the сorrelators is that they are not symmetric in eigenvalues — and thus can not be expressed through Schur polynomials only. Instead, we demonstrate that an expression is possible in terms of Schur derivatives.
This opens a new way to study arbitrary IZ correlators of any order in character expansion.

}

\bigskip

\bigskip

\section{Introduction}

Matrix models \cite{MAMO} have proved to be a powerful tool for the study of non-perturbative phenomena and hidden integrability in quantum field theory \cite{UFN3}. Still, while much is already known for Hermitian and complex matrices, the understanding of the unitary case \cite{0906.3518} -- which can be closer to Yang-Mills dynamics -- remains relatively poor. The reason is already the non-trivial Haar measure for the unitary group, but more important is the relevance of non-trivial actions, of which the typical example is the IZ one \cite{IZ}. Despite satisfying the Duistermaat-Heckman consistency \cite{DH} between the action and the measure, what makes the theory exactly solvable, an explicit expression for generic correlators is still not found. Moreover, the existing formulas in particular cases \cite{9209074,Sha,EyFe} long looked rather heavy and distracted people from going deeper into the problem. However, the recent progress \cite{WLZZ,MO} with applications of the IZ character expansion -- i.e. decomposition of IZ integral into a sum over Schur polynomials \cite{Bal,0906.3518}, which looks just slightly different from the Hermitian case -- should give new momentum to these studies. This paper is an attempt to formulate the program of this research and provide the first manifestations that it can be successful and illuminating.\\
\newline
Namely, we explain how to extract correlators from the character expansion -- which is not a fully direct procedure, but we demonstrate that it can be made practical. As a basic example, we show consistency with the known formulas for pair correlators, which were quite difficult to deduce by alternative methods. We approached the possibilities to figure out this problem by solving a system of linear equations, which perfectly works for the simple case of a $2\times2$ matrix. But for $N\geq3$, we end up with fewer necessity conditions. While this approach needs much more attention but it provides an intuition to look for Ward identities, which we explained at the end of this paper. However, the past results \cite{9209074} provide a detailed view of the pair correlators in determinant form, which illustrates the non-symmetrical nature of this. We discovered that such correlators can be decomposed in the sum of Schur polynomial derivatives. Finally, we proposed an ansatz to expand the Schur derivatives to calculate the pair correlators using the previous results. While this particular approach facilitates the appearance of several coefficients in terms of Schur derivatives, a general structure still needs to be found to express the pair correlators entirely in terms of Schur functions. To avoid overloading the presentation with extra technicalities, we postpone the detailed consideration of higher correlators to the future -- but now this does not look as hopeless as before. \\
\newline
This paper is organized as follows. In section 2, we introduce the IZ integral, its determinant and character formula, and show some examples. Next, in section 3, we introduce the IZ correlators and demonstrate its consistency using the past results in the cases of $N=2$ and $N=3$. In this section, we encountered the possibility of deriving the correlators by solving differential equations. In section 4, we derived the non-vanishing correlators from \cite{9209074} and established a general formula for the ansatz to calculate any pair correlator in Schur polynomial derivatives. Finally, in section 5, we illustrate the possibility of going beyond the pair correlator and provide an intuition for the Ward identities for the IZ model.

\section{Itzykson-Zuber Integral}

Unitary matrix models generally appear in description of gauge theories, and our specific interest, the Itzykson–Zuber integral \cite{IZ}, arises, for example, in the Kazakov–Migdal approach \cite{KM,MAMO} to QCD. They proposed a model of Yang-Mills theory in lattice to describe "induced QCD" where the partition function is
\begin{equation}
    Z = \int \prod_{links} \mathcal{D}U \prod_{sites} \mathcal{D} X e^{-S}
\end{equation}
where $U$ is an unitary and $X$ is a Hermitian matrix with the action
\begin{equation}
    S = N \sum_{sites} \tr (V(X)) - N\sum_{links} \tr(XUYU^{-1})
\end{equation}
The solvability of the Kazakov-Migdal model is due to the applicability of Duistermaat-Heckman approach \cite{DH,UFN3} to the
IZ integral \cite{IZ} over the $N\times N$ unitary matrix ($U U^\dagger = U^\dagger U = I$)
\begin{equation}
    I[X,Y] = \int_{N \times N} e^{\tr (XUYU^\dagger )}[dU]
    \label{1}
\end{equation}
which can be explicitly calculated for any $N$.
In fact such unitary-matrix integrals play a role in many branches of theoretical physics and catch attention of mathematicians as well.
They naturally appear in lattice gauge theory and provide an essential tool for dealing with non-perturbative quantum field theory.\\
\newline
In the integral (\ref{1}), $[dU]$ is a Haar measure, and X ,Y are Hermitian matrices ($X = X^\dagger$). It is the non-trivial Haar measure  that complicates the analysis. Still there are at least two ways to express the exact answer. \\

It depends on matrices $X$ and $Y$ only through their eigenvalues $x_i$ and $y_i$, and the original IZ formula is \cite{IZ}:
\begin{equation}
    I[X,Y] = \frac{{\rm det}\ e^{x_a y_b}}{\Delta (X)\Delta (Y)} =  c_N \sum_P (-)^P\frac{e^{\sum_k x_k y_{P(k)}}}{\Delta(X)\Delta(Y)}
    \label{2}
\end{equation}
where $\Delta (X)$ and $\Delta (Y)$ are the Vandermonde determinant defined as $$ \Delta (X) = \prod_{i<j} (x_i - x_j) \ \ \ \  \Delta (Y) = \prod_{i<j} (y_i - y_j).$$
This formula can be considered as the Duistermaat-Heckman one.\\

Alternatively, in the character expansion formalism, the integral (\ref{1}) is a sum of product of Schur polynomials of corresponding matrix $X$ and $Y$ over  integer partitions. Namely  \cite{Bal,0906.3518},
\begin{equation}
    I[X,Y] =  \sum_{R} \frac{S_R\{ \delta_{k,1} \} S_R[X] S_R[Y]}{S_R[N]}
    \label{3}
\end{equation}
where the sum goes over all the Young diagrams $R$.

To understand the origin of this summation,
we can begin from exponential expansion for  the $2\times 2$ matrix $\Psi = {\rm diag}(x_1,x_2)$:
$$  e^{\tr \Psi} = 1 + x_1+x_2 + \frac{(x_1+x_2)^2}{2} + ..... =
    $$
    \begin{equation}
    = 1+\underbrace{x_1+x_2}_{S_{\ydiagram{1}}}  + \underbrace{\frac{1}{2}}_{S_{\ydiagram{2}}\{ \delta_{k,1} \}} \underbrace{ \frac{(x_1^2 +x_2^2) + (x_1+x_2)^2}{2} }_{S_{\ydiagram{2}}}+  \underbrace{\frac{1}{2}}_{S_{\ydiagram{1,1}}\{ \delta_{k,1} \}}
\underbrace{ \frac{-(x_1^2 +x_2^2) + (x_1+x_2)^2}
{2}}_{S_{\ydiagram{1,1}}} + ....
    \label{4}
\end{equation}
and note that these are the first terms of
expansion, which actually involves the Schur polynomials
\begin{equation}
    e^{\tr \Psi} = \sum_R S_R\{ \delta_{k,1} \}S_R\{ \tr \Psi^k\}
    \label{5}
\end{equation}
and is known as  Cauchy formula \cite{0906.3518,Cauchy}.
It is actually true for matrix $\Psi$ of any size $N\times N$  in the fundamental representation of $sl_N$.

We can also interpret these Schur polynomials as characters, i.e.
the traces of a matrix in non-trivial representations.
Namely the trace of a matrix ${\cal M}$ in representation $R$ is expressed through traces of its powers $M^k$ in the fundamental representations:
\begin{equation}
    {\rm Tr}_R \, {\cal M}  = S_R \{ \tr \ M^k \}
    \label{6565}
\end{equation}

Now we can proceed to integrals.
The main one will be the orthogonality property for unitary matrices ${\cal U}$ in any representation $R$:
\begin{equation}
    \int {\cal U}_{IJ} {\cal U}_{KL}^\dagger\, [dU] = \frac{\delta_{IK} \delta_{LJ}}{D_R} = \frac{\delta_{IK} \delta_{LJ}}{S_R[N]}
    \label{6767}
\end{equation}
Combining the properties (\ref{5})-(\ref{6565}), we can now express the  IZ integral (\ref{1}) as
$$
   I[X,Y] = \int {e^{\tr (XUYU^\dagger)} [dU]} \overset{(\ref{5})}{=} \sum_R S_R \{ \delta_{k,1} \} \int S_R \{ \tr X^kUY^kU^\dagger\} [dU]\overset{(\ref{6565})}{=} \sum_R S_R \{ \delta_{k,1} \} \int {\rm Tr}_R ({\cal XUYU}^\dagger) [dU] \overset{(\ref{6767})}{=}
$$
\begin{equation}
  \overset{(\ref{6767})}{=} \sum_R S_R \{ \delta_{k,1} \} \frac{1}{S_R[N]} {\rm Tr}  {\cal X} \  {\rm Tr} {\cal Y}   \overset{(\ref{6565})}{=}  \sum_R \frac{S_R\{ \delta_{k,1} \}S_R\{ \tr X^k\} S_R\{ \tr Y^k\}}{S_R[N]}
    \label{6}
\end{equation}
and (\ref{3}) is abbreviated form of this formula. \\

Now, we want to extend this relation between the integral the Schur polynomial forms
not for IZ {\it per se}, but for arbitrary correlators with the IZ measure.
Namely, one can define the n-point IZ correlator  as follows
\begin{equation}
\braket{U_{i_1j_1} U_{k_1l_1}^\dagger...U_{i_nj_n} U_{k_nl_n}^\dagger} = \int U_{i_1j_1} U_{k_1l_1}^\dagger...U_{i_nj_n} U_{k_nl_n}^\dagger e^{{\rm tr}(XUYU^\dagger)} [dU]
\label{61}
\end{equation}
where $U$'s are just the ordinary $N\times N$ unitary matrices, i.e. belong to the fundamental representation.
In this article, we mainly focus on the pair correlator and try to shed more light on the search for  the character expansion
of (\ref{61}).

Before we continue with the correlator formalism, it would be beneficial to examine some straightforward examples of integral calculations from both expressions (\ref{1}) and (\ref{3}). This could provide us with a general picture of how the expansion order in the integral is related to the sum of the Young diagrams.

\subsection{An example of the IZ integral for $N=2$}

To calculate the integral (\ref{3}), we can sum up to any order of the diagram, but to show the equivalence between the diagram and the expansion order, we proceed with our calculation by first looking at the several basic properties of the Schur polynomial. This can be written in the p-variable and x-variable, but in this paper, we will use the coordinate variable $x,y$ in further calculations. \\
\newline
If we write the Schur polynomials in power sum variables $(p_k)$, then the constant $S_R\{ \delta_{k,1} \}$ will be the coefficient of the term which contains only $p_1$.
$$ \text{For} \ \ S_{\ydiagram{1}} = p_1: \ \ S_{\ydiagram{1}} \{\delta_{k,1} \} = 1 \ \ \text{and} \ \ S_{\ydiagram{2}} = \frac{p_1^2}{2}+\frac{p_2}{2} \ \ \text{and} \ \  S_{\ydiagram{1,1}} = \frac{p_1^2}{2}-\frac{p_2}{2}:   \ \   S_{\ydiagram{2}} \{\delta_{k,1} \} = S_{\ydiagram{1,1}} \{\delta_{k,1} \} = \frac{1}{2} $$
To calculate the second constant in the formula $S_R[N]$, we need to express the Schur polynomials in $p-$variables and put N in all the places of $p$. For example,
$$S_{\ydiagram{1}} [2] = p_1 = 2, \ \ S_{\ydiagram{2}} [2] = \frac{p_1^2}{2}+\frac{p_2}{2} = 3 \ \ \text{and} \ \ S_{\ydiagram{1,1}} [2] = \frac{p_1^2}{2}-\frac{p_2}{2} = 1 $$
As we will delve into the calculation of higher order integrals and correlators in the future, listing some of these coefficients will be useful.
\begin{table}[H]
		\centering
		\begin{tabular}[h!]{c|c|c}
			Young diagram & $S_R\{ \delta_{k,1} \}$ & $S_R [2]$   \\ &&\\
			\hline &&\\
			$\ydiagram{1}$ & $ 1 $ & 2  \\ &&\\
			\hline &&\\
			$\ydiagram{2}$ & $\frac{1}{2}$ & 3  \\ &&\\
			\hline &&\\
			$\ydiagram{1,1}$ & $ \frac{1}{2}$ & 1 \\ &&\\
			\hline && \\
			$\ydiagram{3}$ & $\frac{1}{6}$ & 4 \\ &&\\
                \hline && \\
                $\ydiagram{2,1}$ & $\frac{1}{3}$ & 2 \\ &&\\
                \hline &&\\
                $\ydiagram{4}$ & $\frac{1}{24}$ & 5 \\ &&\\
                \hline &&\\
                $\ydiagram{3,1}$ & $\frac{1}{8}$ & 3 \\ &&\\
                \hline &&\\
                $\ydiagram{2,2}$ & $\frac{1}{12}$ & 1 \\
		\end{tabular}
  \label{T1}
  \caption{  Coefficients in the Schur expression of the integral (\ref{1})}
	\end{table}
For simplicity, we calculate the IZ integral in power sum polynomials by summing up all the diagrams for all the partitions of 1 and 2. For the blank diagram, we will get a 1 in the sum.
$$I = 1 + \frac{S_{\ydiagram{1}} \{\delta_{k,1} \} S_{\ydiagram{1}} [X] S_{\ydiagram{1}} [Y]}{S_{\ydiagram{1}} [2]} + \frac{S_{\ydiagram{2}} \{\delta_{k,1} \} S_{\ydiagram{2}} [X] S_{\ydiagram{2}} [Y]}{S_{\ydiagram{2}} [2]} + \frac{S_{\ydiagram{1,1}} \{\delta_{k,1} \} S_{\ydiagram{1,1}} [X] S_{\ydiagram{1,1}} [Y]}{S_{\ydiagram{1,1}} [2]}$$
It gives
$$ I = 1 + \frac{1}{12} \left(p_1[y]^2 \left(2 p_1[x]^2-p_2[x]\right)+6 p_1[y] p_1[x]-p_2[y] \left(p_1[x]^2-2 p_2[x]\right)\right)$$
This gives the final expression of the integral
\begin{equation}
    I = \frac{1}{6} \left(3 x_1 x_2 y_1 y_2+3 \left(x_1+x_2\right) \left(y_1+y_2\right)+\left(x_1^2+x_2 x_1+x_2^2\right) \left(y_1^2+y_2 y_1+y_2^2\right)+6\right)
    \label{7}
\end{equation}
On the other side, to calculate the integral (\ref{1}), we can use different parametrization for the unitary matrix. We are using the Euler angle parametrization for $SU(2)$. This gives
$$ XUYU^\dagger = \begin{pmatrix}
x_1 & 0\\
0 & x_2
\end{pmatrix} \begin{pmatrix}
e^{i\phi}cos\theta & e^{i\psi}sin\theta  \\
-e^{-i\psi}sin\theta & e^{-i\phi}cos\theta
\end{pmatrix}\begin{pmatrix}
y_1 & 0\\
0 & y_2
\end{pmatrix} \begin{pmatrix}
e^{-i\phi}cos\theta & -e^{i\psi}sin\theta\\
e^{-i\psi}sin\theta & e^{i\phi}cos\theta
\end{pmatrix} = $$
\bigskip
$$ = \left(
\begin{array}{cc}
 x_1 y_2 \sin ^2(\theta )+x_1 y_1 \cos ^2(\theta ) & x_1 y_2 \sin (\theta ) \cos (\theta ) e^{\text{i$\psi $}+\text{i$\phi $}}-x_1 y_1 \sin (\theta ) \cos (\theta )
   e^{\text{i$\psi $}+\text{i$\phi $}} \\ \\
 x_2 y_2 \sin (\theta ) \cos (\theta ) e^{-\text{i$\psi $}-\text{i$\phi $}}-x_2 y_1 \sin (\theta ) \cos (\theta ) e^{-\text{i$\psi $}-\text{i$\phi $}} & x_2 y_1 \sin
   ^2(\theta )+x_2 y_2 \cos ^2(\theta ) \\
\end{array}
\right)$$

\bigskip
\noindent
The trace is
$$\tr(XUYU^\dagger) = x_2 \left(y_1 \sin ^2(\theta )+y_2 \cos ^2(\theta )\right)+x_1 \left(y_2 \sin ^2(\theta )+y_1 \cos ^2(\theta )\right)$$
\newline
To calculate the Haar measure $dU$ for the parametrization of $N\times N$ matrix with elements $\lambda_i$, it gives
\begin{equation}
    dU = J(\lambda_1,\lambda_2,.....,\lambda_N) d\lambda_1 d\lambda_2.....d\lambda_N
\end{equation}
In our case of the Euler angle parametrization, the Jacobian is $sin(2\theta)$, and the Haar measure with the normalization condition is
\begin{equation}
    \int dU = \int sin(2\theta) d\theta d\phi d\psi = 1
    \label{8}
\end{equation}
Now we can evaluate the integral (\ref{1}) by following
\begin{equation}
    I = \int_0^{2\pi}d\phi\int_0^{2\pi}d\psi \int_0^{\pi/2} e^{x_2 \left(y_1 \sin ^2(\theta )+y_2 \cos ^2(\theta )\right)+x_1 \left(y_2 \sin ^2(\theta )+y_1 \cos ^2(\theta )\right)} sin(2\theta) d\theta =
    \label{9} \nn
\end{equation}
\begin{equation}
    = \frac{e^{x_2 y_1+x_1 y_2}-e^{x_1 y_1+x_2 y_2}}{ \left(x_2-x_1\right) \left(y_1-y_2\right)}
    \label{10}
\end{equation}
Here, we can compare the exponential expansion of the expression (\ref{10}) with the sum of the Young diagram. If we expand both of the exponential of (\ref{10}) up to order 2, then it will be equivalent to the evaluation of calculating the integral in Schur polynomial of $S_{\ydiagram{1}}$. If we expand up to order 3, then the integral will be equivalent to calculating the sum $ S_{\emptyset} + S_{\ydiagram{1}}+S_{\ydiagram{2}}+S_{\ydiagram{1,1}}$. For expansion of 4th order, the integral will be $ S_{\emptyset} + S_{\ydiagram{1}}+S_{\ydiagram{2}}+S_{\ydiagram{1,1}}+ S_{\ydiagram{3}}+ S_{\ydiagram{2,1}}+ S_{\ydiagram{1,1,1}}$ and so on. So in our example.
\begin{equation}
    e^{x_2 y_1+x_1 y_2} = \underbrace{1 + (x_2 y_1+x_1 y_2) + \frac{(x_2 y_1+x_1 y_2)^2}{2!} }_{{\text{for the partition up to 2}}}+...
    \label{11}
\end{equation}
\begin{equation}
    e^{x_1 y_1+x_2 y_2} = \underbrace{1 + (x_1 y_1+x_2 y_2) + \frac{(x_1 y_1+x_2 y_2)^2}{2!} }_{{\text{for the partition up to 2}}}+...
    \label{12}
\end{equation}
Putting (\ref{11}) and (\ref{12}) into (\ref{10}), we get exactly the expression (\ref{7}).
\section{Itzykson-Zuber correlators}
So far, we have seen the IZ integral in three forms. The integral form itself, the determinant form, and the Schur polynomial form. To build up these three formalisms for the IZ correlators, we start from the general integral form (\ref{61}). Taking such integrals for large-$N$ is technically difficult.
Fortunately, a determinant formula  was suggested in \cite{9209074} for the pair correlators:
\begin{equation}
    \braket{|U_{ij}|^2} = V_N \sum_{P} (-)^P \frac{exp\left( \sum_k x_k y_{P(k)}\right)}{ \Delta (X) \Delta (Y) } A_{ij}^{(P)} [X,Y]
    \label{mor1}
\end{equation}
where $A_{ij}^{(P)} [X,Y]$ are some coefficients and the main subject of our study.
In  \cite{9209074} they were defined as
$$
    A_{ij}^{P} =  \delta_{jP(i)} \left( 1 - \sum_{k \neq i} \frac{1}{(x_i - x_k) (y_{P(i)} - y_{P(k)})} + \sum_{k\neq l\neq i \ k\neq i}  \frac{1}{(x_i - x_k)(x_i - x_l) (y_{P(i)} - y_{P(l)})} - ....\right) + $$
\begin{equation}
    + \left( 1- \delta_{jP(i)} \right) \left( - \frac{1}{(x_i - x_{P^{-1}(j)}) (y_j - y_{P(i)})} + \sum_{l \neq P^{-1}(j) \neq i \ l \neq i}  \frac{1}{(x_i - x_{P^{-1}(j)}) (x_i - x_{l}) (y_j - y_{P(i)}) (y_j - y_{P(l)})} \right)
    \label{mor2}
\end{equation}
Later, B.Eynard \cite{EyFe} generalized this formula in a simpler form as
  \begin{equation}
       \braket{U_{ij} U_{ji}} = Res_{x\to x_i} Res_{y \to y_j} \Biggl \langle tr \left( \frac{1}{x-X} U\frac{1}{y-Y}U^\dagger \right) \biggr \rangle  = 1 - det \left( 1 - \frac{1}{x-X} e^{xy} \frac{1}{y-Y} e^{-xy} \right)
  \end{equation}
Now, using (\ref{mor1}) and (\ref{mor2}), we can calculate all the non-vanishing correlators for any $N$.
\subsection{The case of N=2}
To demonstrate how the formula (\ref{mor1}) works, we look at the simple example of a $2\times2$ matrix. We show the consistency of the integral formula, this determinant formula and derive an intermediate expression in terms of the differentiated Schur polynomials, which also matches with the previous result. For that, let's begin writing (\ref{mor1}) and (\ref{mor2}) in the case of $N=2$. For convenience, we will use the following notation: $X_{mn} = x_m - x_n $
\begin{equation}
    \braket{U_{11}U_{11}^\dagger} = \braket{U_{22}U_{22}^\dagger} = \left( \frac{e^{x_1 y_1 + x_2 y_2}}{X_{12} Y_{12}} \left( 1- \frac{1}{X_{12} Y_{12}} \right)  + \frac{e^{x_1 y_2 + x_2 y_1}}{(X_{12} Y_{12})^2} \right)
\end{equation}
\begin{equation}
    \braket{U_{12}U_{21}^\dagger} = \braket{U_{21}U_{12}^\dagger} =  \frac{e^{x_1 y_1 + x_2 y_2}}{(X_{12} Y_{12})^2} - \frac{e^{x_1 y_2 + x_2 y_1}}{X_{12} Y_{12}} + \left( 1+\frac{1}{X_{12} Y_{12}} \right)
\end{equation}
Now, expanding these exponentials, we can get the correlators up to any grading. For example, the correlators with grading 2 are
\begin{equation}
    \braket{U_{11}U_{11}^\dagger} = \braket{U_{22}U_{22}^\dagger} = \frac{1}{6} \left(x_1 \left(2 y_1+y_2\right)+x_2 \left(y_1+2 y_2\right)+3\right)
    \label{2222}
\end{equation}
\begin{equation}
\braket{U_{12}U_{21}^\dagger} = \braket{U_{21}U_{12}^\dagger} =  \frac{1}{6} \left(x_2 \left(2 y_1+y_2\right)+x_1 \left(y_1+2 y_2\right)+3\right)
\label{2333}
\end{equation}
We get similar results by taking the integral. The pair correlators in integral form
\begin{equation}
\braket{U_{ij} U_{kl}^\dagger} = \int U_{ij}U_{kl}^\dagger e^{tr(XUYU^\dagger)} dU
\label{15}
\end{equation}
As we have previously calculated the integral, now we can easily get some non-vanishing correlators using this
\begin{equation}
     \braket{U_{11} U_{11}^\dagger} = \frac{e^{x_1 y_1+x_2 y_2} \left(\left(x_1-x_2\right) \left(y_1-y_2\right)+e^{-\left(\left(x_1-x_2\right)
   \left(y_1-y_2\right)\right)}-1\right)}{\left(x_1-x_2\right){}^2 \left(y_1-y_2\right){}^2} = \braket{U_{22} U_{22}^\dagger}
   \label{16}
\end{equation}
\begin{equation}
     \braket{U_{12} U_{21}^\dagger} = \frac{e^{x_2 y_1+x_1 y_2} \left(-\left(\left(x_1-x_2\right) \left(y_1-y_2\right)\right)+e^{\left(x_1-x_2\right)
   \left(y_1-y_2\right)}-1\right)}{\left(x_1-x_2\right){}^2 \left(y_1-y_2\right){}^2} = \braket{U_{21} U_{12}^\dagger}
   \label{17}
\end{equation}
\begin{equation}
    \braket{U_{11} U_{22} U_{11}^\dagger U_{22}^\dagger} = \frac{e^{x_1 y_1+x_2 y_2} \left(\left(x_1-x_2\right) \left(y_1-y_2\right) \left(\left(x_1-x_2\right) \left(y_1-y_2\right)-2\right)-2 e^{-\left(\left(x_1-x_2\right)
   \left(y_1-y_2\right)\right)}+2\right)}{\left(x_1-x_2\right){}^3 \left(y_1-y_2\right){}^3}
   \label{18}
\end{equation}
{\scriptsize\begin{equation} \braket{U_{11} U_{12} U_{11}^\dagger U_{21}^\dagger} = \frac{e^{x_2 y_1+x_1 y_2} \left(x_1 \left(y_1-y_2\right) \left(e^{\left(x_1-x_2\right) \left(y_1-y_2\right)}+1\right)-x_2 \left(y_1-y_2\right)
   \left(e^{\left(x_1-x_2\right) \left(y_1-y_2\right)}+1\right)-2 e^{\left(x_1-x_2\right) \left(y_1-y_2\right)}+2\right)}{\left(x_1-x_2\right){}^3
       \left(y_1-y_2\right){}^3} \nn
       \label{19}
       \end{equation}}
{\scriptsize\begin{equation}\braket{U_{11} U_{22} U_{12} U_{11}^\dagger U_{22}^\dagger U_{21}^\dagger} =\frac{2 e^{x_2 y_1+x_1 y_2} \left(-\left(\left(x_1-x_2\right) \left(y_1-y_2\right)\right)-3\right)+e^{x_1 y_1+x_2 y_2} \left(\left(x_1-x_2\right) \left(y_1-y_2\right) \left(\left(x_1-x_2\right)
   \left(y_1-y_2\right)-4\right)+6\right)}{\left(x_1-x_2\right){}^4 \left(y_1-y_2\right){}^4} \nn
   \label{20}
   \end{equation}}

\noindent
Expression (\ref{16}) and (\ref{17}) perfectly match with (\ref{2222}) and (\ref{2333}).\\
\newline
Generally, differentiating the integral should lead us to the correlator formalism. As we already know the determinant form of the integral, we now can differentiate both sides of (\ref{2}) by $X_{ij}$:
\begin{equation}
\frac{\partial}{\partial X_{ij}} \left( \int_{N \times N} e^{tr (XUYU^\dagger )}[dU] \right) = \frac{\partial}{\partial X_{ij}} \left(  c_N \sum_P (-)^P\frac{e^{\sum_k x_k y_{P(k)}}}{\Delta(X)\Delta(Y)} \right) = \sum_k \frac{\partial x_k}{\partial X_{ij}}\frac{\partial (I[X,Y])}{\partial x_k}
\label{21}
\end{equation}
This expression contains two derivatives. One is the derivative of the eigenvalues by the matrix element itself, and another is the derivative of the integral by the eigenvalue. For now, let us make derivatives of the eigenvalues by the matrix element and keep it as is. For the choice of diagonal matrix X, the differentiation will be obvious, but for non-diagonal matrix, it needs a more accurate setup. So, the differentiation of the integral by the eigenvalues of $X$ gives\\
\newline
\begin{equation}
     \frac{\partial (IZ)}{\partial x_k} = \frac{\partial}{\partial x_k} \left(  c_N \sum_P (-)^P\frac{e^{\sum_k x_k y_{P(k)}}}{\Delta(X)\Delta(Y)} \right) = \frac{c_N \sum_P (-)^P}{\Delta (Y)} \frac{\Delta(X) y_{P(k)}e^{\sum_k x_k y_{P(k)}} - e^{\sum_k x_k y_{P(k)} }\Delta(X) \sum_{j\neq k} \frac{1}{x_k - x_j}}{(\Delta (X))^2}
     \nn
\end{equation}
Which simplifies
\begin{equation}
    \frac{\partial (IZ)}{\partial x_k} = c_N \sum_P (-)^P \frac{e^{\sum_m x_m y_{P(m)}}(y_{P(k)} - \sum_{l\neq k} \frac{1}{x_k - x_l})}{\Delta(X) \Delta(Y)}
    \label{22}
\end{equation}
Now, simplifying the left-hand side of the equation (\ref{21}), we get
\begin{equation}
    \frac{\partial}{\partial X_{ij}} \left( \int_{N \times N} e^{tr (XUYU^\dagger )}[dU] \right) =  \left<(UYU^\dagger)_{ij}\right> =  \sum_{mn} \braket{U_{im}U_{nj}^\dagger}Y_{mn}
\end{equation}
Finally, combining everything
\begin{equation}
     \sum_{mn} \braket{U_{im}U_{nj}^\dagger}Y_{mn} =  \sum_k  \frac{\partial (I)}{\partial x_k} \frac{\partial x_k}{\partial X_{ij}} =  c_N \sum_k \sum_P (-)^P \frac{e^{\sum_m x_m y_{P(m)}}(y_{P(k)} - \sum_{l\neq k} \frac{1}{x_k - x_l})}{\Delta(X) \Delta(Y)} \frac{\partial x_k}{\partial X_{ij}}
     \label{23}
\end{equation}
It gives the sum of correlators multiplied by the matrix elements $Y_{mn}$. To demonstrate how the integral formula is consistent with the determinant formula, we can look at the simplest example of $i,j=1$.
Then we can calculate the derivatives of the eigenvalues $x_k$ with respect to the variation of matrix $X$ --
at the point where the matrix ix diagonal.
Then
\be
x_{1,2} = \frac{X_{11}+X_{22} \pm \sqrt{(X_{11}-X_{22})^2 + 4X_{12}X_{21}}}{2}
\ee
and
\be
\frac{\p x_1}{\p X_{11}} = \frac{1}{2}\left(1 + \frac{X_{11}-X_{22}}{\sqrt{(X_{11}-X_{22})^2 + 4X_{12}X_{21}}}\right)
\ \stackrel{X_{12}=X_{21}=0}{\longrightarrow} \ 1 , \nn \\
\frac{\p x_2}{\p X_{11}} = \frac{1}{2}\left(1 - \frac{X_{11}-X_{22}}{\sqrt{(X_{11}-X_{22})^2 + 4X_{12}X_{21}}}\right)
\ \stackrel{X_{12}=X_{21}=0}{\longrightarrow} \  0
\ee

Then from the equation (\ref{23}):
\be
    \braket{U_{11}U_{11}^\dagger}y_{1} +  \braket{U_{12}U_{21}^\dagger}y_{2}  =
  \frac{1}{\Delta(X)\Delta(Y)}\left( e^{x_1y_1 + x_2y_2} \left( y_1 - \frac{1}{x_1 -x_2} \right) - e^{x_1y_2 +x_2y_1} \left( y_2 - \frac{1}{x_1 -x_2} \right) \right) =
\nn \\
=      \frac{x_1 \left(y_1 e^{x_1 y_1+x_2 y_2}-y_2 e^{x_2 y_1+x_1 y_2}\right)+e^{x_2 y_1+x_1
   y_2}-e^{x_1 y_1+x_2 y_2}+x_2 \left(y_2 e^{x_2 y_1+x_1 y_2}-y_1 e^{x_1 y_1+x_2
   y_2}\right)}{\left(x_1-x_2\right){}^2 \left(y_1-y_2\right)} \ \ \ \ \ \ \
   \label{25}
\ee
This is in full accirdance with (\ref{16}) and (\ref{17}), but now we derived this {\it combined} relation
directly from the {\it properties} of the integral, without calculating it.
Also, one can easily check the remarkable property of the correlator for this simple example:
the sum
\begin{equation}
     \braket{U_{11}U_{11}^\dagger} +  \braket{U_{12}U_{21}^\dagger} = I[X,Y]
     \label{inti}
\end{equation}
equals the IZ integral itself.

\bigskip

 Now, in search for the correlator in Schur form, we can simply differentiate the Schur version of the integral (\ref{3}) by the matrix element $X_{ij}$ and $Y_{kl}$. This gives us the following two equations
\begin{equation}
    \sum_{mn} \braket{U_{im}U_{nj}^\dagger} Y_{mn} = \sum_R \frac{S_R\{ \delta_{k,1} \}}{S_R[N]}\frac{\partial S_R[X]}{\partial X_{ij}}  S_R[Y]
    \label{26}
\end{equation}
\begin{equation}
        \sum_{pq} \braket{U_{kp}U_{ql}^\dagger} X_{pq} = \sum_R \frac{S_R\{ \delta_{k,1} \}}{S_R[N]} S_R[X] \frac{ \partial S_R[Y]}{\partial Y_{kl}}
        \label{27}
\end{equation}
Now, for our example of $i,j,k,l=1$, a simple algebra can provide the following two equations on the correlators and the Schur form, with some free $x$ and $y$:
\begin{equation}
    \braket{U_{11}U_{11}^\dagger} y_1 + \braket{U_{12}U_{21}^\dagger} y_2 =  \sum_R \frac{S_R\{ \delta_{k,1} \}}{S_R[N]}\frac{\partial S_R[X]}{\partial x_1}  S_R[Y]
    \label{28}
\end{equation}
\begin{equation}
     \braket{U_{11}U_{11}^\dagger} x_1 + \braket{U_{12}U_{21}^\dagger} x_2 =  \sum_R \frac{S_R\{ \delta_{k,1} \}}{S_R[N]}\frac{\partial S_R[Y]}{\partial y_1}  S_R[X]
     \label{29}
\end{equation}
But there is a remark on the two following equations. To demonstrate the consistency, we notice that both of the correlators (\ref{16}) and (\ref{17}) contain not just the exponential term like in the integral but also a combination of $x,y$ with grading $2$ multiplied. On the other hand, the Schur form in the RHS has a different structure, where there is no free $x,y$ outside of the Schur function. As a result, the overall grading of LHS and RHS has been differently constructed. When we expand the exponential terms in a same order for all the coefficients (including $x$ and $y$) and calculate the total RHS of (\ref{28}) and (\ref{29}), it will not be exactly equal to RHS terms by terms but some terms of a fixed grading will match. As we increase the expansion order of the exponential in LHS and Young diagram in RHS, this matching terms or grading will also increase. We have calculated up to several expansion orders to see how grading is related to this overall consistency match.
\begin{table}[H]
		\centering
		\begin{tabular}[h!]{c|c|c}
			Expansion order & Sum up to & common term   \\
			\hline
			$2$ & $ \ydiagram{1} $ & $\frac{1}{2} \left(y_1+y_2\right)$  \\
			\hline
			$3$ & $\ydiagram{1,1}$ & $\frac{1}{6} \left(3 x_2 y_1 y_2+\left(2 x_1+x_2\right) \left(y_1^2+y_2 y_1+y_2^2\right)+3 \left(y_1+y_2\right)\right)$ \\
			\hline
			$4$ & $ \ydiagram{2,1}$ & $\frac{1}{24} (12 x_2 y_1 y_2+4 x_2 \left(2 x_1+x_2\right) y_1 \left(y_1+y_2\right) y_2+4 \left(2 x_1+x_2\right) $ \\ &  & $\left(y_1^2+y_2 y_1+y_2^2\right)+\left(3
   x_1^2+2 x_2 x_1+x_2^2\right) \left(y_1^3+y_2 y_1^2+y_2^2 y_1+y_2^3\right)+12 \left(y_1+y_2\right))$ \\
			\hline
			$5$ & $\ydiagram{2,2}$ & $\frac{1}{120} (20 x_1 x_2^2 y_1^2 y_2^2+60 x_2 y_1 y_2+20 x_2 \left(2 x_1+x_2\right) y_1 \left(y_1+y_2\right) $ \\ &  & $ y_2+5 x_2 \left(3 x_1^2+2 x_2 x_1+x_2^2\right)
   y_1 \left(y_1^2+y_2 y_1+y_2^2\right) $ \\ &  & $ y_2+20 \left(2 x_1+x_2\right) \left(y_1^2+y_2 y_1+y_2^2\right)+5 \left(3 x_1^2+2 x_2 x_1+x_2^2\right)\left(y_1^3+y_2
   y_1^2+y_2^2 y_1+y_2^3\right) $ \\ &  & $ +\left(4 x_1^3+3 x_2 x_1^2+2 x_2^2 x_1+x_2^3\right) \left(y_1^4+y_2 y_1^3+y_2^2 y_1^2+y_2^3 y_1+y_2^4\right)+60
   \left(y_1+y_2\right))$ \\
                \hline

		\end{tabular}
  \label{T1}
  \caption{  Comparison of the expansion order and common term in both sides of (\ref{28})}
	\end{table}
But actually, if we expand the same exponential in different orders depending on what is multiplied with it, then the problem has been solved, and we get a perfect grading match in both sides. \\
\newline
Now, by solving equations (\ref{28}) and (\ref{29}), we can get two unique non-vanishing correlators in the following form.
\begin{equation}
    \braket{U_{11}U_{11}^\dagger} = \sum_R \frac{S_R\{ \delta_{k,1} \}}{S_R[N]} \left( \frac{\frac{\partial S_R[X]}{\partial x_1}  S_R[Y] x_2 - \frac{\partial S_R[Y]}{\partial y_1}  S_R[X] y_2 }{x_2 y_1 - x_1 y_2} \right)
    \label{30}
\end{equation}
\begin{equation}
     \braket{U_{12}U_{21}^\dagger} = \sum_R \frac{S_R\{ \delta_{k,1} \}}{S_R[N]} \left( \frac{\frac{\partial S_R[X]}{\partial x_1}  S_R[Y] x_1 - \frac{\partial S_R[Y]}{\partial y_1}  S_R[X] y_1 }{y_2 x_1 - x_2 y_1} \right)
     \label{31}
\end{equation}
We can look at some examples of the correlators here in the same way as (\ref{28}). Again, we have to expand the exponential, and the story of grading match appears. As it needs to be matched with the terms in both LHS and RHS, we again to need to expand the exponential in different orders depending on what is multiplied with it. But for this simple example, we have checked that if we expand the exponential in the same order despite what is multiplied with it then the following relation holds.
\begin{table}[H]
		\centering
		\begin{tabular}[h!]{c|c|c}
			Expansion order & Young diagram & Maximum grading   \\
			\hline
			$2$ & $\ydiagram{1}$ & $0$  \\
			\hline
            $3$ & $\ydiagram{1,1}$ & $2$  \\
			\hline
			$4$ & $\ydiagram{2,1}$ & $4$ \\
			\hline
            $5$ & $\ydiagram{2,2}$ & $6$ \\
			\hline
            $6$ & $\ydiagram{3,2}$ & $8$ \\
			\hline
            $7$ & $\ydiagram{4,2}$ & $10$ \\
			\hline
            $8$ & $\ydiagram{7,2}$ & $12$ \\
			\hline
            $\vdots$ & $\vdots$ & $\vdots$ \\
            	\hline
			
		\end{tabular}
  \label{T5}
  \caption{Relations between the expansion order, Young diagram and maximum grading in (\ref{30})}
	\end{table}
Moreover, a careful observation of (\ref{30}) and (\ref{31}) reveal that we can write these two expressions in a determinant form as well.
\begin{equation}
    \braket{U_{11} U_{11}^\dagger} = \sum_R \frac{S_R\{ \delta_{k,1} \}}{S_R[N]} \frac{{\rm det} \begin{pmatrix}
\frac{\partial S_R[X]}{\partial x_1} S_R[Y] & \frac{\partial S_R[Y]}{\partial y_1} S_R[X]  \\
y_2 & x_2
\end{pmatrix}}{det \begin{pmatrix}
x_2& x_1  \\
y_2 & y_1
\end{pmatrix} }
\end{equation}
\begin{equation}
     \braket{U_{12} U_{21}^\dagger} = \sum_R \frac{S_R\{ \delta_{k,1} \}}{S_R[N]} \frac{{\rm det} \begin{pmatrix}
\frac{\partial S_R[X]}{\partial x_1} S_R[Y] & \frac{\partial S_R[Y]}{\partial y_1} S_R[X]  \\
y_1 & x_1
\end{pmatrix}}{det \begin{pmatrix}
y_2& y_1  \\
x_2 & x_1
\end{pmatrix} }
\end{equation}
As we see that this is an intermediate stage of our goal. The correlator should resemble $x,y$ variables inside the Schur function, and they should not present independently outside. There is a possible attempt we can do is to calculate several terms of (\ref{30}) and (\ref{31}) and look if the overall result can give us some different combinations of Schur or their derivatives or not. We made a primary attempt for the correlator$ \braket{U_{11}U_{11}^\dagger}$ here.

\begin{equation}
    \braket{U_{11}U_{11}^\dagger} = \frac{1}{2}. \underbrace{1} \frac{\partial S_{\ydiagram{1,0}}[X] }{\partial x_1}  + \frac{1}{6} \underbrace{\frac{1}{2}} \left( \frac{\partial S_{\ydiagram{2,0}}[X] }{\partial x_1} \frac{\partial S_{\ydiagram{2,0}}[Y] }{\partial y_1}\right) \underbrace{-\frac{1}{2}}\left( \frac{\partial S_{\ydiagram{1,1}}[X] }{\partial x_1} \frac{\partial S_{\ydiagram{1,1}}[Y] }{\partial y_1}\right) + ....
    \label{32}
\end{equation}
We have calculated up to the diagram $\ydiagram{1,1}$ and found that an additional coefficient appears in front of each term. As this coefficient is different from the known one, continuing this series for a higher diagram and doing the same calculation for other correlators of different $N$ might provide us a general form of the coefficient. A detailed and more accurate description is present in the next section.
\subsection{The case of N=3}
In the previous subsection, we approached calculating the correlators by solving linear equations (\ref{28}) and (\ref{29}). Where (\ref{30}) and (\ref{31}) provide an expression for the unique non-vanishing pair correlator for $2\times2$ matrix. Here, we will extend our approach to a $3\times3$ matrix and look at how the denominator and the coefficients in Schur are changing. For this, we run the dummy indices of (\ref{26}) and (\ref{27}) up to 3 and make the following system of equations:
\begin{equation}
     \braket{U_{11}U_{11}^\dagger} y_1 +  \braket{U_{12}U_{21}^\dagger} y_2 + \braket{U_{13}U_{31}^\dagger} y_3 =  \sum_R \frac{S_R\{ \delta_{k,1} \}}{S_R[N]}\frac{\partial S_R[X]}{\partial x_1}  S_R[Y] = \mathfrak{S}_R^1
     \label{477}
\end{equation}
\begin{equation}
     \braket{U_{21}U_{12}^\dagger} y_1 +  \braket{U_{22}U_{22}^\dagger} y_2 + \braket{U_{23}U_{32}^\dagger} y_3 =  \sum_R \frac{S_R\{ \delta_{k,1} \}}{S_R[N]}\frac{\partial S_R[X]}{\partial x_2}  S_R[Y] =  \mathfrak{S}_R^2
\end{equation}
\begin{equation}
     \braket{U_{31}U_{13}^\dagger} y_1 +  \braket{U_{32}U_{23}^\dagger} y_2 + \braket{U_{33}U_{33}^\dagger} y_3 =  \sum_R \frac{S_R\{ \delta_{k,1} \}}{S_R[N]}\frac{\partial S_R[X]}{\partial x_3}  S_R[Y] =  \mathfrak{S}_R^3
\end{equation}

\begin{equation}
     \braket{U_{11}U_{11}^\dagger} x_1 +  \braket{U_{12}U_{21}^\dagger} x_2 + \braket{U_{13}U_{31}^\dagger} x_3 =  \sum_R \frac{S_R\{ \delta_{k,1} \}}{S_R[N]}\frac{\partial S_R[Y]}{\partial y_1}  S_R[X] = \mathfrak{P}_R^1
\end{equation}
\begin{equation}
     \braket{U_{21}U_{12}^\dagger} x_1 +  \braket{U_{22}U_{22}^\dagger} x_2 + \braket{U_{23}U_{32}^\dagger} x_3 =  \sum_R \frac{S_R\{ \delta_{k,1} \}}{S_R[N]}\frac{\partial S_R[Y]}{\partial y_2}  S_R[X] =  \mathfrak{P}_R^2
\end{equation}
\begin{equation}
     \braket{U_{31}U_{13}^\dagger} x_1 +  \braket{U_{32}U_{23}^\dagger} x_2 + \braket{U_{33}U_{33}^\dagger} x_3 =  \sum_R \frac{S_R\{ \delta_{k,1} \}}{S_R[N]}\frac{\partial S_R[Y]}{\partial y_3}  S_R[X] =  \mathfrak{P}_R^3
     \label{522}
\end{equation}
For now, we have 9 correlators that we want to find and 6 equations.
$$
    \braket{U_{11}U_{11}^\dagger}, \  \braket{U_{22}U_{22}^\dagger}, \  \braket{U_{33}U_{33}^\dagger}, \  \braket{U_{12}U_{21}^\dagger}, \
     \braket{U_{21}U_{12}^\dagger}, \
      \braket{U_{13}U_{31}^\dagger}, \
       \braket{U_{31}U_{13}^\dagger}, \
        \braket{U_{23}U_{32}^\dagger}, \
        \braket{U_{32}U_{23}^\dagger}, \ $$
In the previous subsection we have seen a symmetry between the correlators like $\braket{U_{11}U_{11}^\dagger} = \braket{U_{22}U_{22}^\dagger} $ and $\braket{U_{12}U_{21}^\dagger} = \braket{U_{21}U_{12}^\dagger}$. As the system in this case has more equations than the variables (correlators), we first need to investigate if there is any symmetry between the correlators. But to look for the symmetry, we need to calculate the correlators first using some existing methods. Hopefully, we have equations (\ref{mor1}) and \ref{mor2}, using which we can calculate all the non-vanishing correlators.
\begin{multline}
    \braket{U_{11}U_{11}^\dagger} = \frac{1}{\Delta (X) \Delta (Y)} \bigg(( e^{x_1 y_1 + x_2y_2 + x_3y_3} \left( 1 - \frac{1}{X_{12} Y_{12} }- \frac{1}{X_{13} Y_{13} } + \frac{1}{X_{12} X_{13} Y_{12} Y_{13}} \right) - \\ -
e^{x_1 y_1 + x_2 y_3 + x_3 y_2} \left( 1 - \frac{1}{X_{12} Y_{13}} - \frac{1}{X_{13} Y_{12} } + \frac{1}{X_{12} X_{13} Y_{12} Y_{13}}  \right) - e^{x_1 y_2 + x_2 y_1 + x_3 y_3} \left( -\frac{1}{X_{12} Y_{12}} + \frac{1}{X_{12} X_{13} Y_{12} Y_{13}} \right) + \\ + e^{x_1 y_2 + x_2 y_3 + x_3 y_1} \left( -\frac{1}{X_{13} Y_{12}} + \frac{1}{X_{12} X_{13} Y_{12} Y_{13}} \right) - e^{x_1 y_3 + x_2 y_2 + x_3 y_1} \left( -\frac{1}{X_{13} Y_{13}} + \frac{1}{X_{12} X_{13} Y_{12} Y_{13}} \right) + \\ + e^{x_1 y_3 + x_2 y_1 + x_3 y_2} \left( -\frac{1}{X_{12} Y_{13}} + \frac{1}{X_{12} X_{13} Y_{12} Y_{13}} \right)  \bigg)
\end{multline}

\begin{multline}
    \braket{U_{12}U_{21}^\dagger} = \frac{1}{\Delta (X) \Delta (Y)} \bigg(( e^{x_1 y_1 + x_2y_2 + x_3y_3} \left( \frac{1}{X_{12} Y_{12} }- \frac{1}{X_{12} X_{13} Y_{12} Y_{23}} \right) - \\ -
e^{x_1 y_1 + x_2 y_3 + x_3 y_2} \left( \frac{1}{X_{13} Y_{12} } - \frac{1}{X_{12} X_{13} Y_{12} Y_{23}}  \right) - e^{x_1 y_2 + x_2 y_1 + x_3 y_3} \left( 1+ \frac{1}{X_{12} Y_{12}} - \frac{1}{X_{13} Y_{23}} - \frac{1}{X_{12} X_{13} Y_{12} Y_{23}} \right) + \\ + e^{x_1 y_2 + x_2 y_3 + x_3 y_1} \left(1 - \frac{1}{X_{12} Y_{23}} + \frac{1}{X_{13} Y_{12}} - \frac{1}{X_{12} X_{13} Y_{13} Y_{23}} \right) - e^{x_1 y_3 + x_2 y_2 + x_3 y_1} \left( -\frac{1}{X_{12} Y_{23}} - \frac{1}{X_{12} X_{13} Y_{12} Y_{23}} \right) + \\ + e^{x_1 y_3 + x_2 y_1 + x_3 y_2} \left( -\frac{1}{X_{13} Y_{23}} - \frac{1}{X_{12} X_{13} Y_{12} Y_{23}} \right)  \bigg)
\end{multline}

\begin{multline}
    \braket{U_{13}U_{31}^\dagger} = \frac{1}{\Delta (X) \Delta (Y)} \bigg(( e^{x_1 y_1 + x_2y_2 + x_3y_3} \left( \frac{1}{X_{13} Y_{13} } + \frac{1}{X_{12} X_{13} Y_{13} Y_{23}} \right) - \\ -
e^{x_1 y_1 + x_2 y_3 + x_3 y_2} \left( \frac{1}{X_{12} Y_{13} } + \frac{1}{X_{12} X_{13} Y_{13} Y_{23}}  \right) - e^{x_1 y_2 + x_2 y_1 + x_3 y_3} \left(\frac{1}{X_{13} Y_{23}} + \frac{1}{X_{12} X_{13} Y_{13} Y_{23}} \right) + \\ + e^{x_1 y_2 + x_2 y_3 + x_3 y_1} \left( \frac{1}{X_{12} Y_{23}} + \frac{1}{X_{12} X_{13} Y_{13} Y_{23}} \right) - e^{x_1 y_3 + x_2 y_2 + x_3 y_1} \left(1 + \frac{1}{X_{12} Y_{23}} +\frac{1}{X_{13} Y_{13}} + \frac{1}{X_{12} X_{13} Y_{13} Y_{23}} \right) + \\ + e^{x_1 y_3 + x_2 y_1 + x_3 y_2} \left(1 +\frac{1}{X_{12} Y_{13}}+\frac{1}{X_{13} Y_{23}} + \frac{1}{X_{12} X_{13} Y_{13} Y_{23}} \right)  \bigg)
\end{multline}

\begin{multline}
    \braket{U_{22}U_{22}^\dagger} = \frac{1}{\Delta (X) \Delta (Y)} \bigg(( e^{x_1 y_1 + x_2y_2 + x_3y_3} \left( 1 -  \frac{1}{X_{23} Y_{23}} - \frac{1}{X_{21} Y_{21}} + \frac{1}{X_{23} X_{21} Y_{23} Y_{21}} \right) - \\ -
e^{x_1 y_1 + x_2 y_3 + x_3 y_2} \left( -\frac{1}{X_{23} Y_{23} } + \frac{1}{X_{23} X_{21} Y_{23} Y_{21}}  \right) - e^{x_1 y_2 + x_2 y_1 + x_3 y_3} \left(-\frac{1}{X_{21} Y_{21}} + \frac{1}{X_{23} X_{21} Y_{23} Y_{21}}\right) + \\ + e^{x_1 y_2 + x_2 y_3 + x_3 y_1} \left( -\frac{1}{X_{21} Y_{23}} + \frac{1}{X_{23} X_{21} Y_{23} Y_{21}} \right) - e^{x_1 y_3 + x_2 y_2 + x_3 y_1} \left( 1 - \frac{1}{X_{23} Y_{21}} - \frac{1}{X_{21} Y_{23}} + \frac{1}{X_{23} X_{21} Y_{23} Y_{21}} \right) + \\ + e^{x_1 y_3 + x_2 y_1 + x_3 y_2} \left( - \frac{1}{X_{23} Y_{21}} + \frac{1}{X_{23} X_{21} Y_{23} Y_{21}} \right)  \bigg)
\end{multline}

\begin{multline}
    \braket{U_{33}U_{33}^\dagger} = \frac{1}{\Delta (X) \Delta (Y)} \bigg(( e^{x_1 y_1 + x_2y_2 + x_3y_3} \left( 1 -  \frac{1}{X_{32} Y_{32}} - \frac{1}{X_{31} Y_{31}} + \frac{1}{X_{32} X_{31} Y_{32} Y_{31}} \right) - \\ -
e^{x_1 y_1 + x_2 y_3 + x_3 y_2} \left(- \frac{1}{X_{32} Y_{32} } + \frac{1}{X_{32} X_{31} Y_{32} Y_{31}}  \right) - e^{x_1 y_2 + x_2 y_1 + x_3 y_3} \left(1 - \frac{1}{X_{32} Y_{31}} - \frac{1}{X_{31} Y_{32}} + \frac{1}{X_{32} X_{31} Y_{32} Y_{31}} \right) + \\ + e^{x_1 y_2 + x_2 y_3 + x_3 y_1} \left( -\frac{1}{X_{32} Y_{31}} + \frac{1}{X_{32} X_{31} Y_{32} Y_{31}} \right) - e^{x_1 y_3 + x_2 y_2 + x_3 y_1} \left(-\frac{1}{X_{31} Y_{31}} + \frac{1}{X_{32} X_{31} Y_{32} Y_{31}} \right) + \\ + e^{x_1 y_3 + x_2 y_1 + x_3 y_2} \left( - \frac{1}{X_{31} Y_{32}} + \frac{1}{X_{32} X_{31} Y_{32} Y_{31}} \right)  \bigg)
\end{multline}

\begin{multline}
    \braket{U_{21}U_{12}^\dagger} = \frac{1}{\Delta (X) \Delta (Y)} \bigg(( e^{x_1 y_1 + x_2y_2 + x_3y_3} \left(- \frac{1}{X_{21} Y_{12}} + \frac{1}{X_{21} X_{23} Y_{12} Y_{13}} \right) - \\ -
e^{x_1 y_1 + x_2 y_3 + x_3 y_2} \left(- \frac{1}{X_{21} Y_{13}} - \frac{1}{X_{21} X_{23} Y_{12} Y_{13}}  \right) -e^{x_1 y_2 + x_2 y_1 + x_3 y_3} \left( 1 - \frac{1}{X_{21} Y_{12}} - \frac{1}{X_{23} Y_{13}} + \frac{1}{X_{21} X_{23} Y_{12} Y_{13}} \right) + \\ + e^{x_1 y_2 + x_2 y_3 + x_3 y_1} \left( -\frac{1}{X_{23} Y_{13}} + \frac{1}{X_{21} X_{23} Y_{12} Y_{13}} \right) - e^{x_1 y_3 + x_2 y_2 + x_3 y_1} \left(-\frac{1}{X_{23} Y_{12}} + \frac{1}{X_{21} X_{23} Y_{12} Y_{13}} \right) + \\ + e^{x_1 y_3 + x_2 y_1 + x_3 y_2} \left( 1 - \frac{1}{X_{23} Y_{12}} - \frac{1}{X_{21} Y_{13}} + \frac{1}{X_{21} X_{23} Y_{12} Y_{13}} \right)  \bigg)
\end{multline}

\begin{multline}
    \braket{U_{23}U_{32}^\dagger} = \frac{1}{\Delta (X) \Delta (Y)} \bigg(( e^{x_1 y_1 + x_2y_2 + x_3y_3} \left(- \frac{1}{X_{23} Y_{32}} + \frac{1}{X_{23} X_{21} Y_{32} Y_{31}} \right) - \\ -
e^{x_1 y_1 + x_2 y_3 + x_3 y_2} \left( 1 - \frac{1}{X_{21} Y_{31}} - \frac{1}{X_{23} Y_{32}} + \frac{1}{X_{23} X_{21} Y_{32} Y_{31}} \right) - e^{x_1 y_2 + x_2 y_1 + x_3 y_3} \left( - \frac{1}{X_{23} Y_{31}} + \frac{1}{X_{23} X_{21} Y_{32} Y_{31}} \right) + \\ + e^{x_1 y_2 + x_2 y_3 + x_3 y_1} \left( 1 -\frac{1}{X_{21} Y_{32}} - \frac{1}{X_{23} Y_{31}} + \frac{1}{X_{23} X_{21} Y_{32} Y_{31}} \right) - e^{x_1 y_3 + x_2 y_2 + x_3 y_1} \left(-\frac{1}{X_{21} Y_{32}} + \frac{1}{X_{23} X_{21} Y_{32} Y_{31}} \right) + \\ + e^{x_1 y_3 + x_2 y_1 + x_3 y_2} \left(- \frac{1}{X_{21} Y_{31}} + \frac{1}{X_{23} X_{21} Y_{32} Y_{31}} \right)  \bigg)
\end{multline}

\begin{multline}
    \braket{U_{32}U_{23}^\dagger} = \frac{1}{\Delta (X) \Delta (Y)} \bigg(( e^{x_1 y_1 + x_2y_2 + x_3y_3} \left(- \frac{1}{X_{32} Y_{23}} + \frac{1}{X_{32} X_{31} Y_{23} Y_{21}} \right) - \\ -
e^{x_1 y_1 + x_2 y_3 + x_3 y_2} \left( 1 - \frac{1}{X_{31} Y_{21}} - \frac{1}{X_{32} Y_{23}} +  \frac{1}{X_{32} X_{31} Y_{23} Y_{21}} \right) - e^{x_1 y_2 + x_2 y_1 + x_3 y_3} \left(- \frac{1}{X_{31} Y_{23}} +  \frac{1}{X_{32} X_{31} Y_{23} Y_{21}} \right) + \\ + e^{x_1 y_2 + x_2 y_3 + x_3 y_1} \left(-\frac{1}{X_{31} Y_{21}} - \frac{1}{X_{32} X_{31} Y_{23} Y_{21}} \right) - e^{x_1 y_3 + x_2 y_2 + x_3 y_1} \left(-\frac{1}{X_{32} Y_{21}} + \frac{1}{X_{32} X_{31} Y_{23} Y_{21}} \right) + \\ + e^{x_1 y_3 + x_2 y_1 + x_3 y_2} \left(1 - \frac{1}{X_{31} Y_{23}}- \frac{1}{X_{32} Y_{21}} +  \frac{1}{X_{32} X_{31} Y_{23} Y_{21}} \right)  \bigg)
\end{multline}

\begin{multline}
    \braket{U_{31}U_{13}^\dagger} = \frac{1}{\Delta (X) \Delta (Y)} \bigg(( e^{x_1 y_1 + x_2y_2 + x_3y_3} \left(- \frac{1}{X_{31} Y_{13}} + \frac{1}{X_{31} X_{32} Y_{13} Y_{12}} \right) - \\ -
e^{x_1 y_1 + x_2 y_3 + x_3 y_2} \left( - \frac{1}{X_{31} Y_{12}} + \frac{1}{X_{31} X_{32} Y_{13} Y_{12}} \right) - e^{x_1 y_2 + x_2 y_1 + x_3 y_3} \left( - \frac{1}{X_{32} Y_{13}} +  \frac{1}{X_{31} X_{32} Y_{13} Y_{12}} \right) + \\ + e^{x_1 y_2 + x_2 y_3 + x_3 y_1} \left(1 - \frac{1}{X_{31} Y_{12}} -\frac{1}{X_{32} Y_{13}} - \frac{1}{X_{31} X_{32} Y_{13} Y_{12}} \right) - e^{x_1 y_3 + x_2 y_2 + x_3 y_1} \left( 1 - \frac{1}{X_{31} Y_{13}} - \frac{1}{X_{32} Y_{12}} + \frac{1}{X_{31} X_{32} Y_{13} Y_{12}} \right) + \\ + e^{x_1 y_3 + x_2 y_1 + x_3 y_2} \left( - \frac{1}{X_{32} Y_{12}} +  \frac{1}{X_{31} X_{32} Y_{13} Y_{12}} \right)  \bigg)
\end{multline}
These following expressions of correlators are written in the form of exponential and Vandermonde determinant. By carefully expanding the exponentials, depending on which term is multiplied with it, we can calculate the correlators up to any grading. But for simplicity and in search of symmetry, let's calculate them up to grading 2. We get the following

\begin{equation}
     \braket{U_{11}U_{11}^\dagger}
= \frac{1}{48} \left( 2 x_1 \left(2 y_1+y_2+y_3\right)+x_2 \left(2 y_1+3  y_2+3 y_3\right)
+ x_3(2  y_1+3   y_2+3   y_3) +8\right)
\label{first}
\end{equation}
\begin{equation}
    \braket{U_{22}U_{22}^\dagger}  = \frac{1}{48} \left( x_1 \left(3 y_1+2 y_2+3 y_3\right)+2 x_2 \left(y_1+2 y_2+y_3\right) + x_3 (3 y_1+2 y_2+3 y_3) +8\right)
\end{equation}
\begin{equation}
     \braket{U_{33}U_{33}^\dagger}  = \frac{1}{48} \left(x_1 \left(3 y_1+3 y_2+2 y_3\right)+x_2 \left(3 y_1+3 y_2+2 y_3\right) + 2 x_3 \left(y_1+y_2+2 y_3\right)+8\right)
\end{equation}
\begin{equation}
     \braket{U_{12}U_{21}^\dagger} = \frac{1}{48} \left(2 x_1 \left(y_1+2 y_2+y_3\right)+x_2 \left(3 y_1+2 y_2+3 y_3\right) + x_3 (3 y_1+2 y_2+3 y_3)+8\right)
\end{equation}
\begin{equation}
    \braket{U_{21}U_{12}^\dagger} = \frac{1}{48} \left(x_1 \left(2 y_1+3 y_2+3y_3\right) +2 x_2 \left(2 y_1+y_2+y_3\right)+ x_3 (2 y_1+3 y_2+3 y_3) +8\right)
\end{equation}
\begin{equation}
    \braket{U_{13}U_{31}^\dagger} = \frac{1}{48} \left(2 x_1 \left(y_1+y_2+2 y_3\right)+x_2 \left(3 y_1+3 y_2+2 y_3\right) + x_3 (3 y_1+3 y_2+2 y_3)+8\right)
\end{equation}
\begin{equation}
      \braket{U_{31}U_{13}^\dagger}  = \frac{1}{48} \left(x_1 \left(2 y_1+3 y_2+3y_3\right)+x_2 \left(2 y_1+3 y_2+3y_3\right) + 2 x_3 \left(2 y_1+y_2+y_3\right)+8\right)
\end{equation}
\begin{equation}
    \braket{U_{23}U_{32}^\dagger} = \frac{1}{48} \left(x_1 \left(3 y_1+3 y_2+2 y_3\right) +2 x_2 \left(y_1+y_2+2 y_3\right)+x_3 (3 y_1+3 y_2+2 y_3)+8\right)
\end{equation}
\begin{equation}
     \braket{U_{32}U_{23}^\dagger} = \frac{1}{48} \left(x_1 \left(3 y_1+2 y_2+3 y_3\right)+x_2 \left(3 y_1+2 y_2+3 y_3\right) + 2 x_3 \left(y_1+2 y_2+y_3\right)+8\right)
     \label{last}
\end{equation}

Now, these expressions reveal that unlike in the case of $N=2$, these correlators are not equal to each other term by term. Instead, the following symmetry transformations relate them together. \\
\newline
For the diagonal terms
\begin{equation}
    \braket{U_{11}U_{11}} \xleftrightarrow[\text{\small $y_1 \leftrightarrow y_2$}]{\text{\small $x_1 \leftrightarrow x_2$}} \braket{U_{22}U_{22}} \xleftrightarrow[\text{\small $y_2 \leftrightarrow y_3$}]{\text{\small $x_2 \leftrightarrow x_3$}} \braket{U_{33}U_{33}}
\end{equation}
For the off-diagonal terms
\begin{multline}
    \braket{U_{12}U_{21}} \xleftrightarrow[\text{\small $y_1 \leftrightarrow y_2$}]{\text{\small $x_1 \leftrightarrow x_2$}} \braket{U_{21}U_{12}}; \ \braket{U_{13}U_{31}} \xleftrightarrow[\text{\small $y_1 \leftrightarrow y_3$}]{\text{\small $x_1 \leftrightarrow x_3$}} \braket{U_{31}U_{13}}; \ \braket{U_{23}U_{32}} \xleftrightarrow[\text{\small $y_2 \leftrightarrow y_3$}]{\text{\small $x_2 \leftrightarrow x_3$}} \braket{U_{32}U_{23}};
\end{multline}
For off-diagonal mixed terms
\begin{equation}
    \braket{U_{13}U_{31}} \xleftrightarrow[\text{\small $y_2 \leftrightarrow y_3$}]{\text{\small $x_2 \leftrightarrow x_3$}} \braket{U_{12}U_{21}}; \ \braket{U_{13}U_{31}} \xleftrightarrow[\text{\small $y_1 \leftrightarrow y_2$}]{\text{\small $x_1 \leftrightarrow x_2$}} \braket{U_{23}U_{32}};  \ \braket{U_{12}U_{21}} \xleftrightarrow[\text{\small $y_1 \leftrightarrow y_3$}]{\text{\small $x_1 \leftrightarrow x_3$}} \braket{U_{32}U_{23}};
\end{equation}
We also observe a different setting of (\ref{inti}) in this case of $N=3$, which are
\begin{equation}
    \braket{U_{11}U_{11}^\dagger} + \braket{U_{12}U_{21}^\dagger} + \braket{U_{13}U_{31}^\dagger} + \braket{U_{21}U_{12}^\dagger} + \braket{U_{22}U_{22}^\dagger} + \braket{U_{23}U_{32}^\dagger} = I[X,Y]
\end{equation}
\begin{equation}
    \braket{U_{11}U_{11}^\dagger} + \braket{U_{12}U_{21}^\dagger} + \braket{U_{13}U_{31}^\dagger} + \braket{U_{31}U_{13}^\dagger} + \braket{U_{32}U_{23}^\dagger} + \braket{U_{33}U_{33}^\dagger} = I[X,Y]
\end{equation}
\begin{equation}
     \braket{U_{21}U_{12}^\dagger} + \braket{U_{22}U_{22}^\dagger} + \braket{U_{23}U_{32}^\dagger} + \braket{U_{31}U_{13}^\dagger} + \braket{U_{32}U_{23}^\dagger} + \braket{U_{33}U_{33}^\dagger} = I[X,Y]
\end{equation}
As there is no equality between the correlators, we will not be able to simplify the system. So, this approach needs further consideration to extract some similar setup as we did in the case of $N=2$. But still, there is a way to write the correlators in Schur form, a hint of what we got in (\ref{32}). In the next section, we are going to make a Schur expansion of the correlators we can calculate using previous results.

\section{Schur expansion in IZ correlators}

Up to now, we have seen how to calculate the correlator from the integral form in the case of $N=2$ and found an expression in Schur derivatives form with some free $x$ and $y$. We have also calculated all the pair correlators from  formula (\ref{mor2}) for both $N=2$, $N=3$ and found the symmetry relations between them. We see that expressions (\ref{first}) to (\ref{last}) provide the non-symmetric structure of the correlators. Now, it is obvious that a naive combination of just Schur polynomials of $X$ and $Y$ will not appear since they are symmetric. Instead, there will be some non-symmetric functions. In general, there exist many of them, but as we have seen in the previous section, the Schur derivatives evaluate the correlators correctly. So, now, we can look at those expressions and try to make an expansion in Schur serivatives. Our approach is as follows.\\
\newline
We expand the correlators in the following way
\begin{equation}
     \braket{U_{11}U_{11}^\dagger} = t_1 (y) \frac{\partial S_{\ydiagram{1,0}}[X] }{\partial X_{11}} +  t_2 (y) \frac{\partial S_{\ydiagram{2,0}}[X] }{\partial X_{11}} +  t_3 (y) \frac{\partial S_{\ydiagram{1,1}}[X] }{\partial X_{11}}
     \label{77}
\end{equation}
where the coefficients $t_i(y)$ are the combination of some real number and Schur derivatives of $Y$. Then, we compare with the previously calculated expressions and find these coefficients. Our goal is to look for a Schur derivatives structure in these coefficients. For example now if we compared this expansion (\ref{77}) with (\ref{2222}) then we find the coefficients as following
\begin{equation}
    t_1(y) = \frac{1}{2}; \ \  t_2(y) = \frac{1}{12} (2 y_1+y_2); \ \  t_3(y) = \frac{1}{4} y_2
    \label{78}
\end{equation}
We see that these coefficients provide us with some interesting structure
\begin{equation}
    t_1(y) = \frac{1}{2} \frac{\partial S_{\ydiagram{1,0}}[Y]}{\partial Y_{11}}; \ \  t_2(y) = \frac{1}{12} \frac{\partial S_{\ydiagram{2,0}}[Y]}{\partial Y_{11}}; \ \  t_3(y) = \frac{1}{4} \frac{\partial S_{\ydiagram{1,1}}[Y]}{\partial Y_{11}}
\end{equation}
As we result, we finally get
\begin{equation}
    \braket{U_{11}U_{11}^\dagger} = \frac{1}{2} \frac{\partial S_{\ydiagram{1,0}}[X]}{\partial X_{11}} \frac{\partial S_{\ydiagram{1,0}}[Y] }{\partial Y_{11}} + \frac{1}{12} \frac{\partial S_{\ydiagram{2,0}}[X]}{\partial X_{11}} \frac{\partial S_{\ydiagram{2,0}}[Y] }{\partial Y_{11}} +  \frac{1}{4} \frac{\partial S_{\ydiagram{1,1}}[X]}{\partial X_{11}} \frac{\partial S_{\ydiagram{1,1}}[Y] }{\partial Y_{11}}
\end{equation}
Now we can do the same for (\ref{2333}) and calculate the coefficients. We get
\begin{equation}
   \braket{U_{12}U_{21}^\dagger} = \frac{1}{2} \frac{\partial S_{\ydiagram{1,0}}[X]}{\partial X_{11}} \frac{\partial S_{\ydiagram{1,0}}[Y] }{\partial Y_{22}} + \frac{1}{12} \frac{\partial S_{\ydiagram{2,0}}[X]}{\partial X_{11}} \frac{\partial S_{\ydiagram{2,0}}[Y] }{\partial Y_{22}} +  \frac{1}{4} \frac{\partial S_{\ydiagram{1,1}}[X]}{\partial X_{11}} \frac{\partial S_{\ydiagram{1,1}}[Y] }{\partial Y_{22}}
\end{equation}
This method of Schur derivatives expansion now allows us to express (\ref{first})-(\ref{last}) in a similar fashion.
\begin{equation}
     \braket{U_{11}U_{11}^\dagger} = \frac{1}{6} \frac{\partial S_{\ydiagram{1,0}}[X]}{\partial X_{11}} \frac{\partial S_{\ydiagram{1,0}}[Y] }{\partial Y_{11}} + \frac{1}{48} \frac{\partial S_{\ydiagram{2,0}}[X]}{\partial X_{11}} \frac{\partial S_{\ydiagram{2,0}}[Y] }{\partial Y_{11}} +  \frac{1}{24} \frac{\partial S_{\ydiagram{1,1}}[X]}{\partial X_{11}} \frac{\partial S_{\ydiagram{1,1}}[Y] }{\partial Y_{11}}
\end{equation}
\begin{equation}
     \braket{U_{22}U_{22}^\dagger} = \frac{1}{6} \frac{\partial S_{\ydiagram{1,0}}[X]}{\partial X_{22}} \frac{\partial S_{\ydiagram{1,0}}[Y] }{\partial Y_{22}} + \frac{1}{48} \frac{\partial S_{\ydiagram{2,0}}[X]}{\partial X_{22}} \frac{\partial S_{\ydiagram{2,0}}[Y] }{\partial Y_{22}} +  \frac{1}{24} \frac{\partial S_{\ydiagram{1,1}}[X]}{\partial X_{22}} \frac{\partial S_{\ydiagram{1,1}}[Y] }{\partial Y_{22}}
\end{equation}
\begin{equation}
     \braket{U_{33}U_{33}^\dagger} = \frac{1}{6} \frac{\partial S_{\ydiagram{1,0}}[X]}{\partial X_{33}} \frac{\partial S_{\ydiagram{1,0}}[Y] }{\partial Y_{33}} + \frac{1}{48} \frac{\partial S_{\ydiagram{2,0}}[X]}{\partial X_{33}} \frac{\partial S_{\ydiagram{2,0}}[Y] }{\partial Y_{33}} +  \frac{1}{24} \frac{\partial S_{\ydiagram{1,1}}[X]}{\partial X_{33}} \frac{\partial S_{\ydiagram{1,1}}[Y] }{\partial Y_{33}}
\end{equation}
\begin{equation}
     \braket{U_{12}U_{21}^\dagger} = \frac{1}{6} \frac{\partial S_{\ydiagram{1,0}}[X]}{\partial X_{11}} \frac{\partial S_{\ydiagram{1,0}}[Y] }{\partial Y_{22}} + \frac{1}{48} \frac{\partial S_{\ydiagram{2,0}}[X]}{\partial X_{11}} \frac{\partial S_{\ydiagram{2,0}}[Y] }{\partial Y_{22}} +  \frac{1}{24} \frac{\partial S_{\ydiagram{1,1}}[X]}{\partial X_{11}} \frac{\partial S_{\ydiagram{1,1}}[Y] }{\partial Y_{22}}
\end{equation}
\begin{equation}
     \braket{U_{21}U_{12}^\dagger} = \frac{1}{6} \frac{\partial S_{\ydiagram{1,0}}[X]}{\partial X_{22}} \frac{\partial S_{\ydiagram{1,0}}[Y] }{\partial Y_{11}} + \frac{1}{48} \frac{\partial S_{\ydiagram{2,0}}[X]}{\partial X_{22}} \frac{\partial S_{\ydiagram{2,0}}[Y] }{\partial Y_{11}} +  \frac{1}{24} \frac{\partial S_{\ydiagram{1,1}}[X]}{\partial X_{22}} \frac{\partial S_{\ydiagram{1,1}}[Y] }{\partial Y_{11}}
\end{equation}
\begin{equation}
     \braket{U_{13}U_{31}^\dagger} = \frac{1}{6} \frac{\partial S_{\ydiagram{1,0}}[X]}{\partial X_{11}} \frac{\partial S_{\ydiagram{1,0}}[Y] }{\partial Y_{33}} + \frac{1}{48} \frac{\partial S_{\ydiagram{2,0}}[X]}{\partial X_{11}} \frac{\partial S_{\ydiagram{2,0}}[Y] }{\partial Y_{33}} +  \frac{1}{24} \frac{\partial S_{\ydiagram{1,1}}[X]}{\partial X_{11}} \frac{\partial S_{\ydiagram{1,1}}[Y] }{\partial Y_{33}}
\end{equation}
\begin{equation}
     \braket{U_{31}U_{13}^\dagger} = \frac{1}{6} \frac{\partial S_{\ydiagram{1,0}}[X]}{\partial X_{33}} \frac{\partial S_{\ydiagram{1,0}}[Y] }{\partial Y_{11}} + \frac{1}{48} \frac{\partial S_{\ydiagram{2,0}}[X]}{\partial X_{33}} \frac{\partial S_{\ydiagram{2,0}}[Y] }{\partial Y_{11}} +  \frac{1}{24} \frac{\partial S_{\ydiagram{1,1}}[X]}{\partial X_{33}} \frac{\partial S_{\ydiagram{1,1}}[Y] }{\partial Y_{11}}
\end{equation}
\begin{equation}
     \braket{U_{23}U_{32}^\dagger} = \frac{1}{6} \frac{\partial S_{\ydiagram{1,0}}[X]}{\partial X_{22}} \frac{\partial S_{\ydiagram{1,0}}[Y] }{\partial Y_{33}} + \frac{1}{48} \frac{\partial S_{\ydiagram{2,0}}[X]}{\partial X_{22}} \frac{\partial S_{\ydiagram{2,0}}[Y] }{\partial Y_{33}} +  \frac{1}{24} \frac{\partial S_{\ydiagram{1,1}}[X]}{\partial X_{22}} \frac{\partial S_{\ydiagram{1,1}}[Y] }{\partial Y_{33}}
\end{equation}
\begin{equation}
     \braket{U_{32}U_{23}^\dagger} = \frac{1}{6} \frac{\partial S_{\ydiagram{1,0}}[X]}{\partial X_{33}} \frac{\partial S_{\ydiagram{1,0}}[Y] }{\partial Y_{22}} + \frac{1}{48} \frac{\partial S_{\ydiagram{2,0}}[X]}{\partial X_{33}} \frac{\partial S_{\ydiagram{2,0}}[Y] }{\partial Y_{22}} +  \frac{1}{24} \frac{\partial S_{\ydiagram{1,1}}[X]}{\partial X_{33}} \frac{\partial S_{\ydiagram{1,1}}[Y] }{\partial Y_{22}}
\end{equation}
We have seen that the real number stays the same in all the correlators and the same structure of Schur differential remains the same for both $N=2$ and $N=3$ but up to the  diagram $\ydiagram{1,1}$.
\begin{equation}
     \braket{U_{ik}U_{lj}^\dagger} = \sum_{R=\ydiagram{1}}^{\ydiagram{1,1}} \mathcal{C_R}  \frac{\partial S_{R}[X]}{\partial X_{ij}} \frac{\partial S_{R}[Y] }{\partial Y_{kl}}
     \label{909}
\end{equation}
But if we move to the next and even other higher partitions, this structure (\ref{909}) does not hold anymore. To understand the structures of higher partitions, we first have to expand them correctly. More preciously, we need to look for an ansatz that will expand the Schur derivatives in such a way that it provides the exact expression of IZ correlators calculated from  (\ref{mor1}) and (\ref{mor2}). We have found such ansatz which solves the problem correctly and provides a perfect match with the result we find from the past results.\\
\newline
For future use, we denote the coefficients in the following way.
$$
t_{n,R}^{N}(Y) - \text{coefficients of first derivatives}
$$
$$
\tilde{t}_{n,R}^{N} (Y)  - \text{coefficients of second derivatives}
$$
Here, $N$ is the size of the matrix, $R$ is the diagram of the Schur derivatives next to it, and $n$ is the first partition in lexicographic order. Then, for the next several partitions, the following ansatz will hold.
\begin{equation}
     \braket{U_{11}U_{11}^\dagger}|_3 = t_{3, (3,0)}^2 (y) \frac{\partial S_{\ydiagram{3,0}}[X] }{\partial X_{11}} +  t_{3, (2,1)}^2 (y) \frac{\partial S_{\ydiagram{2,1}}[X] }{\partial X_{11}} +  \tilde{t}_{3, (2,1)}^2 (y) x_1 \frac{\partial^2 S_{\ydiagram{2,1}}[X] }{\partial X_{11}^2}
\end{equation}
\begin{equation}
     \braket{U_{11}U_{11}^\dagger}|_4 = t_{4, (4,0)}^2 (y) \frac{\partial S_{\ydiagram{4,0}}[X] }{\partial X_{11}} +  t_{4, (3,1)}^2 (y) \frac{\partial S_{\ydiagram{3,1}}[X] }{\partial X_{11}} +  t_{4, (2,2)}^2 (y) \frac{\partial S_{\ydiagram{2,2}}[X] }{\partial X_{11}} +   \tilde{t}_{4, (2,2)}^2 (y) x_1 \frac{\partial^2 S_{\ydiagram{2,2}}[X] }{\partial X_{11}^2}
\end{equation}
\begin{multline}
     \braket{U_{11}U_{11}^\dagger}|_5 = t_{5, (5,0)}^2 (y) \frac{\partial S_{\ydiagram{5,0}}[X] }{\partial X_{11}} +  t_{5, (4,1)}^2 (y) \frac{\partial S_{\ydiagram{4,1}}[X] }{\partial X_{11}} +  t_{5,(3,2)}^2 (y) \frac{\partial S_{\ydiagram{3,2}}[X] }{\partial X_{11}} + \tilde{t}_{5, (4,1)}^2 (y) x_1 \frac{\partial^2 S_{\ydiagram{4,1}}[X] }{\partial X_{11}^2} + \\  \tilde{t}_{5, (3,2)}^2 (y) x_1 \frac{\partial^2 S_{\ydiagram{3,2}}[X] }{\partial X_{11}^2}
\end{multline}
\begin{multline}
         \braket{U_{11}U_{11}^\dagger}|_6 = t_{6, (6,0)}^2 (y) \frac{\partial S_{\ydiagram{6,0}}[X] }{\partial X_{11}} +  t_{6, (5,1)}^2 (y) \frac{\partial S_{\ydiagram{5,1}}[X] }{\partial X_{11}} +  t_{6, (4,2)}^2  (y) \frac{\partial S_{\ydiagram{4,2}}[X] }{\partial X_{11}} + t_{6, (3,3)}^2  (y) \frac{\partial S_{\ydiagram{3,3}}[X] }{\partial X_{11}}  + \\  \tilde{t}_{6, (5,1)}^2 (y) x_1 \frac{\partial^2 S_{\ydiagram{5,1}}[X] }{\partial X_{11}^2} +  \tilde{t}_{6, (4,2)}^2 (y) x_1 \frac{\partial^2 S_{\ydiagram{4,2}}[X] }{\partial X_{11}^2}
\end{multline}
Now, for the convenience, let's just write the partition instead of Young diagrams
\begin{multline}
         \braket{U_{11}U_{11}^\dagger}|_7 = t_{7, (7,0)}^2 (y) \frac{\partial S_{(7,0)}[X] }{\partial X_{11}} +  t_{7,(6,1)}^2 (y) \frac{\partial S_{(6,1)}[X] }{\partial X_{11}} +  t_{7, (5,2)}^2 (y) \frac{\partial S_{5,2}[X] }{\partial X_{11}} + t_{7, (4,3)}^2 (y) \frac{\partial S_{(4,3)}[X] }{\partial X_{11}} + \\ \tilde{t}_{7, (6,1)}^2 (y) x_1 \frac{\partial^2 S_{(6,1)}[X] }{\partial X_{11}^2} +  \tilde{t}_{7,(5,2)}^2 (y) x_1 \frac{\partial^2 S_{(5,2)}[X] }{\partial X_{11}^2} +  \tilde{t}_{7, (4,3)}^2 (y) x_1 \frac{\partial^2 S_{(4,3)}[X] }{\partial X_{11}^2}
\end{multline}
\begin{multline}
         \braket{U_{11}U_{11}^\dagger}|_8 = t_{8, (8,0)}^2 (y) \frac{\partial S_{(8,0)}[X]}{\partial X_{11}} +  t_{8, (7,1)}^2 (y) \frac{\partial S_{(7,1)}[X] }{\partial X_{11}} +  t_{8, (6,2)}^2 (y) \frac{\partial S_{(6,2)}[X] }{\partial X_{11}} + t_{8, (5,3)}^2 (y) \frac{\partial S_{(5,3)}[X] }{\partial X_{11}} + \\ t_{8, (4,4)}^2 (y) \frac{\partial S_{(4,4)}[X] }{\partial X_{11}} +  \tilde{t}_{8, (7,1)}^2 (y) x_1 \frac{\partial^2 S_{(7,1)}[X] }{\partial X_{11}^2} + \tilde{t}_{8, (6,2)}^2 (y) x_1 \frac{\partial^2 S_{(6,2)}[X] }{\partial X_{11}^2} +  \tilde{t}_{8, (5,3)}^2 (y) x_1 \frac{\partial^2 S_{(5,3)}[X] }{\partial X_{11}^2}
\end{multline}
In the case of $N=3$, the same idea in the ansatz provides the correct expression for the correlators.
\begin{multline}
     \braket{U_{11}U_{11}^\dagger}|_3 = t_{3, (3,0)}^3 (y) \frac{\partial S_{\ydiagram{3,0}}[X] }{\partial X_{11}} +  t_{3, (2,1)}^3 (y) \frac{\partial S_{\ydiagram{2,1}}[X] }{\partial X_{11}} + t_{3, (1,1,1)}^3 (y) \frac{\partial S_{\ydiagram{1,1,1}}[X] }{\partial X_{11}} +  \\ \tilde{t}_{3, (2,1)}^3 (y) x_1 \frac{\partial^2 S_{\ydiagram{2,1}}[X] }{\partial X_{11}^2} +   \tilde{t}_{3, (1,1,1)}^3 (y) x_1 \frac{\partial^2 S_{\ydiagram{1,1,1}}[X] }{\partial X_{11}^2}
\end{multline}
\begin{multline}
   \braket{U_{11}U_{11}^\dagger}|_4 = t_{4, (4,0)}^3 (y) \frac{\partial S_{\ydiagram{4,0}}[X] }{\partial X_{11}} +  t_{4, (3,1)}^3 (y) \frac{\partial S_{\ydiagram{3,1}}[X] }{\partial X_{11}} +  t_{4, (2,2)}^3 (y) \frac{\partial S_{\ydiagram{2,2}}[X] }{\partial X_{11}} + t_{4, (2,1,1)}^3 (y) \frac{\partial S_{\ydiagram{2,1,1}}[X] }{\partial X_{11}} + \\ \tilde{t}_{4,(3,1)}^3 (y) x_1 \frac{\partial^2 S_{\ydiagram{3,1}}[X] }{\partial X_{11}^2} + \tilde{t}_{4, (2,2)}^2 (y) x_1 \frac{\partial^2 S_{\ydiagram{2,2}}[X] }{\partial X_{11}^2}
\end{multline}
\begin{multline}
         \braket{U_{11}U_{11}^\dagger}|_5 = t_{5, (5,0)}^3 (y) \frac{\partial S_{\ydiagram{5,0}}[X] }{\partial X_{11}} +  t_{5, (4,1)}^3 (y) \frac{\partial S_{\ydiagram{4,1}}[X] }{\partial X_{11}} +  t_{5,(3,2)}^3 (y) \frac{\partial S_{\ydiagram{3,2}}[X] }{\partial X_{11}}  +  t_{5, (3,1,1)}^3 (y) \frac{\partial S_{\ydiagram{3,1,1}}[X] }{\partial X_{11}} +  \\ t_{5, (2,2,1)}^3 (y) \frac{\partial S_{\ydiagram{2,2,1}}[X] }{\partial X_{11}}  +   \tilde{t}_{5, (4,1)}^3 (y) x_1 \frac{\partial^2 S_{\ydiagram{4,1}}[X] }{\partial X_{11}^2} + \tilde{t}_{5, (3,2)}^3 (y) x_1 \frac{\partial^2 S_{\ydiagram{3,2}}[X] }{\partial X_{11}^2} + \tilde{t}_{5, (3,1,1)}^3 (y) x_1 \frac{\partial^2 S_{\ydiagram{3,1,1}}[X] }{\partial X_{11}^2} + \\ \tilde{t}_{5, (2,2,1)}^3 (y) x_1 \frac{\partial^2 S_{\ydiagram{2,2,1}}[X] }{\partial X_{11}^2}
\end{multline}
Now, the rest of the correlators can be found by just changing the $ij$ index, and the $kl$ index will be inside the coefficients where $Y$ lies. To write this ansatz in a general form, let's define the following\\
\newline
\noindent\textbf{Definition:}
Let $\mathcal{P}_{n,N} = \left\{ R \vdash n \;\middle|\; \ell(R) \leq N \right\}$, listed in lexicographic order. Let $R_{\min}$ and $R_{\max}$ denote the first and last elements of this ordered set, respectively. Then we define the subset $\mathcal{G}_{n,N}$ as:
\begin{equation}
    \mathcal{G}_{n,N} =
\begin{cases}
\mathcal{P}_{n,N} \setminus \{R_{\min}\}, & \text{if } n \text{ is odd}, \\
\mathcal{P}_{n,N} \setminus \{R_{\min},\, R_{\max}\}, & \text{if } n \text{ is even}.
\end{cases}
\end{equation}
Then, the ansatz for any pair correlators is in the form of Schur derivatives:

\begin{equation}
\boxed{
      \braket{U_{ik}U_{lj}^\dagger} = \sum_{R \in \mathcal{P}_{n,N}} t_{n,R}^{N}(Y) \frac{\partial S_R}{\partial X_{ij}} + X_{ij} \sum_{R \in \mathcal{G}_{n,N}} \tilde{t}_{n,R}^{N} (Y) \frac{\partial^2 S_R}{\partial X_{ij}^2} }
      \label{102}
\end{equation}
Now, the next task is to find a general form of these coefficients $t_{n,R}^{N}(Y) $ and $\tilde{t}_{n,R}^{N} (Y)$ appearing in the expansion. Only then can we fully write the formula (\ref{102}) in the form of character expansion. For this, we calculate some of the coefficients and find the Schur derivative structure in several cases. But finding the general form is still an open problem.
\subsection{Hunt for the coefficients $t_{n,R}^{N}(Y) $ and $\tilde{t}_{n,R}^{N} (Y)$}
\subsubsection{N=2}
So  for now let's list several $t_{n,R}^{N}(Y) $ and $\tilde{t}_{n,R}^{N} (Y)$ for $N=2$ up to $n=8$.
\begin{equation}
     t_{2,(2,0)}^{2}(Y) = \frac{1}{12} \left(2 y_1+y_2\right) =  \frac{1}{12} \frac{\partial S_{\ydiagram{2,0}}[Y]}{\partial y_1} ; \ \   t_{2,(1,1)}^{2}(Y) = \frac{1}{4} y_2  =  \frac{1}{4} \frac{\partial S_{\ydiagram{1,1}}[Y]}{\partial y_1}
\end{equation}
\begin{multline}
     t_{3,(3,0)}^{2}(Y) = \frac{1}{72} \left(3 y_1^2+2 y_2 y_1+y_2^2\right) =  \frac{1}{72} \frac{\partial S_{\ydiagram{3,0}}[Y]}{\partial y_1} ; \\   t_{3,(2,1)}^{2}(Y) = \frac{1}{18} \left(2 y_2^2+y_1 y_2\right) ; \ \ \tilde{t}_{3,(2,1)}^{2} (Y) = \frac{1}{12} \left(y_1 y_2-y_2^2\right) \\
\end{multline}
\begin{multline}
     t_{4,(4,0)}^{2}(Y) = \frac{1}{480} \left(4 y_1^3+3 y_2 y_1^2+2 y_2^2 y_1+y_2^3\right) =  \frac{1}{480} \frac{\partial S_{\ydiagram{4,0}}[Y]}{\partial y_1} ; \ \   t_{4,(3,1)}^{2}(Y) = \frac{1}{96} y_2\left(3 y_2^2+2 y_1 y_2+y_1^2\right) ; \\  t_{4,(2,2)}^{2}(Y) = \frac{1}{144} \left(-y_2^3+6 y_1
   y_2^2+y_1^2 y_2\right); \ \ \tilde{t}_{4,(3,1)}^{2} (Y) = \frac{1}{72} \left(y_1^2 y_2-y_2^3\right) \\
\end{multline}
\begin{multline}
     t_{5,(5,0)}^{2}(Y) = \frac{1}{3600} (5 y_1^4+4 y_2 y_1^3+3 y_2^2 y_1^2+2 y_2^3 y_1+y_2^4) =  \frac{1}{3600} \frac{\partial S_{\ydiagram{5,0}}[Y]}{\partial y_1} \\  t_{5,(4,1)}^{2}(Y) = \frac{1}{600} y_2 \left(y_1^3+2 y_2 y_1^2+3 y_2^2 y_1+4 y_2^3\right) ; \ \  t_{5,(3,2)}^{2}(Y) =  -\frac{1}{720} y_2 \left(-2 y_1^3+y_2 y_1^2-16 y_2^2 y_1+2 y_2^3\right) \\ \tilde{t}_{5,(4,1)}^{2} (Y) = \frac{1}{1440} (y_2 \left(3 y_1^3+y_2 y_1^2-y_2^2 y_1-3 y_2^3\right); \ \ \tilde{t}_{5,(3,2)}^{2} (Y) = \frac{1}{1440} (y_2 \left(-y_1^3+13 y_2 y_1^2-13 y_2^2
   y_1+y_2^3\right)) \\
\end{multline}
\begin{multline}
     t_{6,(6,0)}^{2}(Y) = \frac{6 y_1^5+5 y_2 y_1^4+4 y_2^2 y_1^3+3 y_2^3 y_1^2+2 y_2^4 y_1+y_2^5}{30240} =  \frac{1}{30240} \frac{\partial S_{\ydiagram{6,0}}[Y]}{\partial y_1}; \\   t_{6,(5,1)}^{2}(Y) = \frac{y_2 \left(y_1^4+2 y_2 y_1^3+3 y_2^2 y_1^2+4 y_2^3 y_1+5 y_2^4\right)}{4320}; \ \  t_{6,(4,2)}^{2}(Y) =  \frac{y_2 \left(y_1^4+4 y_2^2 y_1^2+8 y_2^3 y_1-y_2^4\right)}{1920}; \\ t_{6,(3,3)}^{2}(Y) =  \frac{y_2 \left(y_1^4+8 y_2 y_1^3+60 y_2^2 y_1^2-8 y_2^3 y_1-y_2^4\right)}{17280}; \ \ \tilde{t}_{6,(5,1)}^{2} (Y) = \frac{y_2 \left(2 y_1^4+y_2 y_1^3-y_2^3 y_1-2 y_2^4\right)}{7200}; \\ \tilde{t}_{6,(4,2)}^{2} (Y) = \frac{y_2 \left(-y_1^4+12 y_2 y_1^3-12 y_2^3 y_1+y_2^4\right)}{9600};
\end{multline}
\begin{multline}
     t_{7,(7,0)}^{2}(Y) = \frac{7 y_1^6+6 y_2 y_1^5+5 y_2^2 y_1^4+4 y_2^3 y_1^3+3 y_2^4 y_1^2+2 y_2^5 y_1+y_2^6}{282240} = \frac{1}{282240} \frac{\partial S_{\ydiagram{7,0}}[Y]}{\partial y_1}; \\ t_{7,(6,1)}^{2}(Y) = \frac{6 y_2^6+5 y_1 y_2^5+4 y_1^2 y_2^4+3 y_1^3 y_2^3+2 y_1^4 y_2^2+y_1^5 y_2}{35280}; \ \   t_{7,(5,2)}^{2}(Y) =  \frac{-4 y_2^6+34 y_1 y_2^5+23 y_1^2 y_2^4+12 y_1^3 y_2^3+y_1^4 y_2^2+4 y_1^5 y_2}{50400}; \\ t_{7,(4,3)}^{2}(Y) =  \frac{-y_2^6-9 y_1 y_2^5+102 y_1^2 y_2^4-32 y_1^3 y_2^3+9 y_1^4 y_2^2+y_1^5 y_2}{50400}; \ \  \tilde{t}_{7,(6,1)}^{2} (Y) = \frac{-5 y_2^6-3 y_1 y_2^5-y_1^2 y_2^4+y_1^3 y_2^3+3 y_1^4 y_2^2+5 y_1^5 y_2}{151200}; \\  \tilde{t}_{7,(5,2)}^{2} (Y) = \frac{y_2^6-12 y_1 y_2^5-4 y_1^2 y_2^4+4 y_1^3 y_2^3+12 y_1^4 y_2^2-y_1^5 y_2}{75600} \ \  \tilde{t}_{7,(4,3}^{2} (Y) = \frac{y_2^6+9 y_1 y_2^5-172 y_1^2 y_2^4+172 y_1^3 y_2^3-9 y_1^4 y_2^2-y_1^5 y_2}{302400};
\end{multline}
\begin{multline}
    \\ t_{8,(8,0)}^{2}(Y) = \frac{8 y_1^7+7 y_2 y_1^6+6 y_2^2 y_1^5+5 y_2^3 y_1^4+4 y_2^4 y_1^3+3 y_2^5 y_1^2+2 y_2^6 y_1+y_2^7}{2903040} = \frac{1}{2903040} \frac{\partial S_{\ydiagram{8,0}}[Y]}{\partial y_1}; \\  t_{8,(7,1)}^{2}(Y) = \frac{y_2 \left(y_1^6+2 y_2 y_1^5+3 y_2^2 y_1^4+4 y_2^3 y_1^3+5 y_2^4 y_1^2+6 y_2^5 y_1+7 y_2^6\right)}{322560} ; \\  t_{8,(6,2)}^{2}(Y) = \frac{y_2 \left(5 y_1^6+2 y_2 y_1^5+13 y_2^2 y_1^4+24 y_2^3 y_1^3+35 y_2^4 y_1^2+46 y_2^5 y_1-5 y_2^6\right)}{483840}; \\ t_{8,(5,3)}^{2}(Y) =  \frac{y_2 \left(y_1^6+10 y_2 y_1^5-23 y_2^2 y_1^4+40 y_2^3 y_1^3+103 y_2^4 y_1^2-10 y_2^5 y_1-y_2^6\right)}{345600}; \\ t_{8,(4,4)}^{2}(Y) = \frac{y_2 \left(y_1^6+10 y_2 y_1^5+47 y_2^2 y_1^4+420 y_2^3 y_1^3-47 y_2^4 y_1^2-10 y_2^5 y_1-y_2^6\right)}{2419200}; \\ \tilde{t}_{8,(7,1)}^{2} (Y) = \frac{y_2 \left(3 y_1^6+2 y_2 y_1^5+y_2^2 y_1^4-y_2^4 y_1^2-2 y_2^5 y_1-3 y_2^6\right)}{846720};  \\ \tilde{t}_{8,(6,2)}^{2} (Y) = \frac{y_2 \left(-5 y_1^6+62 y_2 y_1^5+31 y_2^2 y_1^4-31 y_2^4 y_1^2-62 y_2^5 y_1+5 y_2^6\right)}{3386880}; \\ \tilde{t}_{8,(5,3}^{2} (Y) = \frac{y_2 \left(-y_1^6-10 y_2 y_1^5+163 y_2^2 y_1^4-163 y_2^4 y_1^2+10 y_2^5 y_1+y_2^6\right)}{2419200};
\end{multline}

\subsubsection{N=3}
In the case of $N=3$ we calculate the coefficients up to n=4:
\begin{equation}
     t_{2,(2,0)}^{3}(Y) = \frac{1}{48} \left(2 y_1+y_2+y_3\right) = \frac{1}{48} \frac{\partial S_{\ydiagram{2,0}}[Y]}{\partial y_1}; \ \   t_{2,(1,1)}^{3}(Y) =\frac{1}{24} \left(y_2+y_3\right) = \frac{1}{24} \frac{\partial S_{\ydiagram{1,1}}[Y]}{\partial y_1};
\end{equation}
\begin{multline}
     t_{3,(3,0)}^{3}(Y) = \frac{1}{360} \left(3 y_1^2+2 y_2 y_1+2 y_3 y_1+y_2^2+y_3^2+y_2 y_3\right) = \frac{1}{360} \frac{\partial S_{\ydiagram{3,0}}[Y]}{\partial y_1}; \\  t_{3,(2,1)}^{3}(Y) = \frac{1}{144} \left(2 y_2^2+y_1 y_2+2 y_3 y_2+2 y_3^2+y_1 y_3\right); \ \ t_{3,(1,1,1)}^{3}(Y) = \frac{1}{72} \left(-y_2^2+y_1 y_2+2 y_3 y_2-y_3^2+y_1 y_3\right); \\ \tilde{t}_{3,(2,1)}^{3} (Y) = \frac{1}{96} \left(-y_2^2+y_1 y_2-y_3^2+y_1 y_3\right);
\end{multline}
\begin{multline}
     t_{4,(4,0)}^{3}(Y) = \frac{4 y_1^3+3 y_2 y_1^2+3 y_3 y_1^2+2 y_2^2 y_1+2 y_3^2 y_1+2 y_2 y_3 y_1+y_2^3+y_3^3+y_2 y_3^2+y_2^2 y_3}{2880} = \frac{1}{2880} \frac{\partial S_{\ydiagram{4,0}}[Y]}{\partial y_1};  \\  t_{4,(3,1)}^{3}(Y) = \frac{1}{960} \left(3 y_2^3+2 y_1 y_2^2+3 y_3 y_2^2+y_1^2 y_2+3 y_3^2 y_2+2 y_1 y_3 y_2+3 y_3^3+2 y_1 y_3^2+y_1^2 y_3\right); \\ t_{4,(2,2)}^{3}(Y) = \frac{-2 y_2^3+5 y_1 y_2^2+9 y_3 y_2^2+2 y_1^2 y_2+9 y_3^2 y_2-8 y_1 y_3 y_2-2 y_3^3+5 y_1 y_3^2+2 y_1^2 y_3}{1440}; \\ t_{4,(2,1,1)}^{3}(Y) = \frac{1}{576} \left(-y_2^3+y_3 y_2^2+y_1^2 y_2+y_3^2 y_2+10 y_1 y_3 y_2-y_3^3+y_1^2 y_3\right); \\  \tilde{t}_{4,(3,1}^{3} (Y) = \frac{-2 y_2^3-y_3 y_2^2+2 y_1^2 y_2-y_3^2 y_2+2 y_1 y_3 y_2-2 y_3^3+2 y_1^2 y_3}{1440}; \\  \tilde{t}_{4,(2,2}^{3} (Y) = \frac{y_2^3-7 y_3 y_2^2-y_1^2 y_2-7 y_3^2 y_2+14 y_1 y_3 y_2+y_3^3-y_1^2 y_3}{1440}
\end{multline}
Although the first coefficient of each partition is visible from this list but it shows a different structure for other cases. So a general formula for these coefficients in Schur derivatives needs to find, which we postpone for future work.

\section{Towards higher correlators and a hope for Ward identities}
In the previous section, formula (\ref{102}) demonstrates how we should make the Schur expansion to restore the correlators. In order to determine the coefficients $t_{n,R}^{N} (Y)$ and $\tilde{t}_{n,R}^{N} (Y)$, we need to compare our ansatz with  the old formulas (\ref{mor1}) and (\ref{mor2}). But they are available only for pair correlators, and if we want to go beyond this, we need to somehow find another way. While a pair correlator fully in the basis of differentiated Schur polynomials might help but looking for a generating function and moving towards Ward identities will be highly beneficial. But unlike the Gaussian Hermitian models, where the generating function is a function of vector, the generating function for unitary correlators will be a function of matrix. This naturally complicates the overall approach but still appears as an active branch for contemporary research. \\
\newline
In Gaussian Hermitian models \cite{UFN3}, it has been considered the correlator of traces, but in this study, we are considering the correlator of matrix elements. This naturally sparks the rigorous treatment of matrix elements in higher order. To visualize this, we can try to differentiate the IZ integral repeatedly by the matrix element $Y_{kl}$ and $X_{ij}$ and try to extract the correlator from there. If we understand this, we can formulate the pair correlator first then the 4-point correlator, and so on in Schur form. In each stage, we have a correlator and a sum of higher correlators, which equal to a Schur structure in the right hand side.  For example, now differentiating (\ref{26}) again by $Y_{kl}$ provides a pair correlator and a sum of a 4-point correlator in the following form:

\begin{equation}
   \frac{\partial}{\partial Y_{kl}} \left( \sum_{mn} \braket{U_{im}U_{nj}} Y_{mn} \right)=  \braket{U_{ik}U_{lj}^\dagger} + \sum_{m,n}\sum_{pq} \braket{U_{im}Y_{mn}U_{nj}^\dagger U_{kp}X_{pq}U_{ql}^\dagger} = \sum_R \frac{S_R\{ \delta_{k,1} \}}{S_R[N]}\frac{\partial S_R[X]}{\partial X_{ij}} \frac{\partial S_R[Y]}{\partial Y_{kl}}
   \label{33}
\end{equation}
We can look at the simple example of $i,j,k,l=1$ and run the dummy indices up to 2 (in the case of $N=2$).\\
\newline
Then the expression (\ref{33}) turns out to be

$$
    \braket{U_{11}U_{11}^\dagger} + \braket{U_{11}U_{11}^\dagger U_{11}U_{11}^\dagger} x_1 y_1 +  \braket{U_{12}U_{21}^\dagger U_{12}U_{21}^\dagger} x_2 y_2 +  \braket{U_{11}U_{11}^\dagger U_{12}U_{21}^\dagger} (x_1 y_2 + x_2 y_1) =
$$
\be
= \sum_R \frac{S_R\{ \delta_{k,1} \}}{S_R[N]}\frac{\partial S_R[X]}{\partial x_1} \frac{\partial S_R[Y]}{\partial y_1}
    \label{35}
\ee
From this, we want to calculate $\braket{U_{ik}U_{lj}^\dagger}$ completely in the form Schur derivatives. But for that, it's necessary to understand the structure of the sum of these four correlators in Schur form. Or maybe if we can somehow get rid of this sum, then it's possible to express the pair correlator in Schur derivatives form. For this, we can differentiate again until we get such a sum of 4-correlator so that we can subtract from the previous one. So, understanding the formalism of n-th derivatives of the IZ integral is an essential step, which can help us to understand both the pair and higher order correlators. \\
\newline
As we are delving into the hope for an expression for n-point correlators, a slightly different index notation from the above might help. Namely, we now write the equation (\ref{33}) in the following notation:
\begin{equation}
\frac{\partial}{\partial Y_{k_2l_2}} \frac{\partial}{\partial X_{i_2j_2}} \left(\braket{U_{i_1k_1}U_{l_1j_1}^\dagger} + \sum_{m_1,n_1}\sum_{p_1q_1} \braket{U_{i_1m_1}Y_{m_1n_1}U_{n_1j_1}^\dagger U_{k_1p_1}X_{p_1q_1}U_{q_1l_1}^\dagger} \right) =
\end{equation}
$$= \frac{\partial}{\partial Y_{k_2l_2}} \frac{\partial}{\partial X_{i_2j_2}} \left(\sum_R \frac{S_R\{ \delta_{k,1} \}}{S_R[N]}\frac{\partial S_R[X]}{\partial X_{i_1j_1}} \frac{\partial S_R[Y]}{\partial Y_{k_1l_1}}  \right)$$
Let's differentiate step by step. Start with differentiating that expression again by X
\begin{equation}
    \frac{\partial }{\partial X_{i_2 j_2}} \left( \braket{U_{i_1k_1}U_{l_1j_1}^\dagger}\right) = \sum_{m_2 n_2} \braket{U_{i_1k_1}U_{l_1j_1 }^\dagger U_{i_2 m_2}U_{n_2j_2}^\dagger} Y_{m_2n_2}
\end{equation}
\begin{equation}
    \frac{\partial }{\partial X_{i_2 j_2}} \left( \sum_{m_1,n_1}\sum_{p_1q_1} \braket{U_{i_1m_1}U_{n_1j_1}^\dagger U_{k_1p_1}U_{q_1l_1}^\dagger}X_{p_1q_1Y_{m_1n_1}} \right) = \sum_{m_1n_1} \braket{U_{i_1m_1}U_{n_1j_1 }^\dagger U_{i_2 k_1}U_{l_1j_2^\dagger}} Y_{m_1n_1} +
    \label{46}
\end{equation}
$$
+ \sum_{m_2n_2}\sum_{p_1q_1} \sum_{m_1n_1} \braket{U_{i_1m_1}U_{n_1j_1  }^\dagger U_{k_1 p_1}U_{q_1l_1}^\dagger U_{i_2 m_2}U_{n_2j_2}^\dagger} Y_{m_1n_1} X_{p_1q_1}  Y_{m_2n_2}
$$
Now we differentiate (\ref{46}) by $Y_{k_2l_2}$
\begin{equation}
      \frac{\partial }{\partial Y_{k_2 l_2}}\left( \frac{\partial }{\partial X_{i_2 j_2}} \left( \braket{U_{i_1k_1}U_{l_1j_1}^\dagger}\right)\right) = \braket{U_{i_1k_1}U_{l_1j_1 }^\dagger U_{i_2 k_2}U_{l_2j_2}^\dagger} + \sum_{p_2q_2} \sum_{m_2 n_2} \braket{U_{i_1k_1}U_{l_1j_1 }^\dagger U_{i_2 m_2}U_{n_2j_2}^\dagger  U_{k_2 p_2}U_{q_2l_2}^\dagger} Y_{m_2n_2} X_{p_2q_2}
\end{equation}
Now, the second part of the differentiation be
$$
     \frac{\partial }{\partial Y_{k_2 l_2}} \left( \frac{\partial }{\partial X_{i_2 j_2}} \left( \sum_{m_1,n_1}\sum_{p_1q_1} \braket{U_{i_1m_1}U_{n_1j_1}^\dagger U_{k_1p_1}U_{q_1l_1}^\dagger}X_{p_1q_1} Y_{m_1n_1} \right) \right) =
$$
\begin{equation}
\braket{U_{i_1k_2}U_{l_2j_1}^\dagger U_{k_1i_2}U_{j_2l_1}^\dagger} + \sum_{p_2q_2} \sum_{m_1 n_1} \braket{U_{i_1m_1}U_{n_1j_1 }^\dagger U_{k_1 i_2}U_{j_2l_1}^\dagger  U_{k_2 p_2}U_{q_2l_2}^\dagger} Y_{m_1n_1} X_{p_2q_2}
\end{equation}
The third part of the differentiation
$$
\frac{\partial}{\partial Y_{k_2l_2}} \left( \sum_{m_2n_2}\sum_{p_1q_1} \sum_{m_1n_1} \braket{U_{i_1m_1}U_{n_1j_1  }^\dagger U_{k_1 p_1}U_{q_1l_1}^\dagger U_{i_2 m_2}U_{n_2j_2}^\dagger} Y_{m_1n_1} X_{p_1q_1}  Y_{m_2n_2} \right) =
$$

$$    \sum_{p_1q_1} \sum_{m_1 n_1} \braket{U_{i_1m_1}U_{n_1j_1 }^\dagger U_{k_1 p_1}U_{q_1l_1}^\dagger  U_{i_2 k_2}U_{l_2j_2}^\dagger} Y_{m_1n_1} X_{p_1q_1} +  \sum_{p_1q_1} \sum_{m_2 n_2} \braket{U_{i_1k_2}U_{l_2j_1 }^\dagger U_{k_1 p_1}U_{q_1l_1}^\dagger  U_{i_2 k_2}U_{l_2j_2}^\dagger} Y_{m_2n_2} X_{p_1q_1} +$$
\begin{equation}
  + \sum_{p_2q_2} \sum_{m_2n_2}\sum_{p_1q_1} \sum_{m_1n_1} \braket{U_{i_1m_1}U_{n_1j_1 }^\dagger U_{k_1 p_1}U_{q_1l_1}^\dagger  U_{i_2 m_2}U_{n_2j_2}^\dagger  U_{k_2 p_2}U_{q_2l_2}^\dagger} Y_{m_1n_1} X_{p_1q_1}  Y_{m_2n_2} X_{p_2q_2}
\end{equation}
Now, combining all the terms, four times differentiation of the IZ integral
$$
    \frac{\partial}{\partial Y_{k_2l_2}} \frac{\partial}{\partial X_{i_2j_2}} \frac{\partial}{\partial Y_{k_1l_1}} \frac{\partial}{\partial X_{i_1j_1}} \left( IZ \right) = \braket{U_{i_1k_1}U_{l_1j_1 }^\dagger U_{i_2 k_2}U_{l_2j_2}^\dagger} +  \braket{U_{i_1k_2}U_{l_2j_1}^\dagger U_{k_1i_2}U_{j_2l_1}^\dagger}+ $$
$$
+ \sum_{p_2q_2} \sum_{m_2 n_2} \braket{U_{i_1k_1}U_{l_1j_1 }^\dagger U_{i_2 m_2}U_{n_2j_2}^\dagger  U_{k_2 p_2}U_{q_2l_2}^\dagger} Y_{m_2n_2} X_{p_2q_2} + \sum_{p_2q_2} \sum_{m_1 n_1} \braket{U_{i_1m_1}U_{n_1j_1 }^\dagger U_{k_1 i_2}U_{j_2l_1}^\dagger  U_{k_2 p_2}U_{q_2l_2}^\dagger} Y_{m_1n_1} X_{p_2q_2} +
$$
$$    \sum_{p_1q_1} \sum_{m_1 n_1} \braket{U_{i_1m_1}U_{n_1j_1 }^\dagger U_{k_1 p_1}U_{q_1l_1}^\dagger  U_{i_2 k_2}U_{l_2j_2}^\dagger} Y_{m_1n_1} X_{p_1q_1} +  \sum_{p_1q_1} \sum_{m_2 n_2} \braket{U_{i_1k_2}U_{l_2j_1 }^\dagger U_{k_1 p_1}U_{q_1l_1}^\dagger  U_{i_2 k_2}U_{l_2j_2}^\dagger} Y_{m_2n_2} X_{p_1q_1} +$$
\begin{equation}
  + \sum_{p_2q_2} \sum_{m_2n_2}\sum_{p_1q_1} \sum_{m_1n_1} \braket{U_{i_1m_1}U_{n_1j_1 }^\dagger U_{k_1 p_1}U_{q_1l_1}^\dagger  U_{i_2 m_2}U_{n_2j_2}^\dagger  U_{k_2 p_2}U_{q_2l_2}^\dagger} Y_{m_1n_1} X_{p_1q_1}  Y_{m_2n_2} X_{p_2q_2}
\end{equation}
 While making differentiation for another several times will make the expression longer, it's obvious that in every step, there will be a sum of correlators of higher order. For example by differentiating $n$ times will provide the n-correlator + sum of some another n-correlators + a sum of 2n-correlator. We see in this way that the big sum in the LHS is poorly related to RHS with a naive derivative of the Schur polynomials. Looking for Ward identities for the IZ integral might help to visualize all the correlators and symmetries between them.\\
 \newline
 Now, continuing this differentiation will not be very convenient with more and more terms and indices, and a different approach is needed to handle them for higher order differentiation. At this moment, using some diagram techniques described in \cite{DolMor} might be very promising. Again, we keep this diagram technique approach for future study.

\section{Conclusion}

In this paper, we addressed the old problem \cite{9209074,Sha,EyFe,0906.3518}
of evaluating the unitary-matrix correlators with Itzykson-Zuber (IZ) measure. Unitary correlators are an especially important chapter of matrix model theory, connecting it to generic Yang-Mills theory — unfortunately, it is rather difficult and attracts insufficient attention. In this paper, we considered the new possibilities opened by the character expansion of IZ integral through Schur polynomials. Knowledge of these explicit formulas provides some information about correlators, but only partial. $2N$ equations like (\ref{477})-(\ref{522}) connect $N^2$ variables $F_{ij} =\ <U_{ij} U^\dagger_{ji}>$, and are not sufficient to define them for $N>2$ as a solution to the linear algebra problem. However, all these quantities are expressed through just two {\it functions} $F_{ii} =G(x_i|x_1,\ldots,\check x_i, \ldots x_N)$ and $F_{ij} = H(x_i,x_j|  |x_1,\ldots,\check x_i,\ldots,\check x_j, \ldots x_N)$ which are symmetric in the variables, different from $x_i$ and $x_j$. Then, using various ansatz or power expansions, we tried to find an explicit expression. While our general formula for the ansatz (\ref{102}) provides the correlators precisely, we encountered a new issue of finding a general formula for the coefficients that appeared in it. We will address this problem in the subsequent publications. What we already achieved in this paper, we checked that explicit formulas of \cite{9209074}  for $G$ and $H$ are consistent with the character expansion. Then we looked at the possibilities of approaching this problem by solving the system of linear equations. Finally, we suggested an ansatz to calculate the correlators in the form of Schur derivatives using the formulas of \cite{9209074}.
Altogether this provides a substantial support to the old guess (\ref{mor2}) and opens a way to proofs and generalizations.
 \\
\newline
To conclude, it is a necessary step towards finding the full IZ partition function, depending on infinitely many time-variables \cite{UFN3} and reproducing {\it all} the IZ correlators. Building up this expression and expressing it through Schur functions remains an open problem. It is especially interesting to see what kind of {\it superintegrability} \cite{MMsi,siunit} will be reflected in it.

\section*{Acknowledgements}
We are  grateful to Anastasia Oreshina, Yaroslav Drachov, Pavel Suprun, Maxim Reva, and Nikita Tselousov for many insightful discussions about this study. Special thanks are to Aleksandr Popolitov for useful feedback and important comments.
\\
\newline
This work is supported by the RSF grant  24-12-00178.

\end{document}